\documentclass[aps,prl,preprint,groupedaddress,titlepage,nofootinbib,longbibliography]{revtex4-1}

\usepackage[pdftex]{graphicx}
\usepackage{dcolumn}
\usepackage{bm}

\newcommand*{\String}{S}
\newcommand*{\multi}{\cdot}

\begin{document}

\preprint{KYUSHU-RCAPP-2019-01}
\preprint{KYUSHU-HET-195}

\title{Hide and Seek with Massive Fields in Modulus Mediation}


\author{Ken-ichi Okumura}
\email{okumura@phys.kyushu-u.ac.jp}
\affiliation{Research Center for Advanced Particle Physics, Kyushu University,\\
744, Motooka, Nishi-ku, Fukuoka 819-0395, Japan\\ }


\date{\today}

 \begin{abstract}
  We study the modulus mediation of supersymmetry breaking motivated by superstring theory. We show that the renormalization group running of the soft supersymmetry breaking parameters due to the interactions of massive fields is canceled by the threshold corrections at one-loop order, if their mass is given by nonperturbative dynamics controlled by the same modulus that mediates supersymmetry breaking and a sum rule of the modular weights holds for the Yukawa couplings. As an example, we discuss order reduction of lepton flavor violation in the supersymmetric seesaw mechanism, which revives the parameter space already excluded.
 \end{abstract}


\maketitle


\section{Introduction}

Supersymmetry (SUSY) is an attractive candidate for physics beyond the standard model (SM).
It stabilizes the big hierarchy between the Planck scale and the electroweak (EW) scale, which is otherwise vulnerable to corrections from the heavy scale.
Furthermore, the minimal SUSY SM (MSSM) predicts gauge coupling unification and the existence of dark matter protected by $R$ parity.
The observed Higgs boson mass is perfectly consistent with the prediction of MSSM and nonobservation of the SUSY particles at the Large Hadron Collider, although the heavy superpartners generally introduce a little fine-tuning in the EW symmetry breaking.

The phenomenology of low energy SUSY mostly depends on the parameters that softly break SUSY and determine the spectrum of the superpartners.
The supertrace constraint requires these parameters to be generated via the nonrenormalizable K\"ahler potential that couples the visible MSSM sector to the hidden sector where SUSY is spontaneously broken \cite{Nilles:1983ge, Martin:1997ns}.
Phenomenologically, the flavor structure of the soft SUSY breaking parameters is tightly constrained by the flavor and $CP$ violating processes providing hints for the origin of the K\"ahler potential. The gauge mediation \cite{Giudice:1998bp}
 and the anomaly mediation \cite{Randall:1998uk, Giudice:1998xp} are widely studied bottom-up models that readily satisfy the constraint.

The anomaly mediation particularly has a special property dubbed ultraviolet insensitivity, which means the soft SUSY breaking parameters in the low energy effective theory are completely determined by the SUSY preserving couplings at the same renormalization scale irrespective of the ultraviolet theory. This stems from the fact that the SUSY breaking effect is encapsulated in the SUSY regulators or, in other words, the renormalization scale dependence and its analytic continuation to the superspace \cite{Giudice:1997ni, ArkaniHamed:1998kj}. In view of the renormalization group (RG) running of the soft parameters, running due to interactions with massive particles is canceled by the threshold corrections and disappears in low energy theory.
This is useful to evade the flavor-$CP$ constraints because the soft SUSY breaking parameters in the MSSM are functions of the gauge and Yukawa couplings at the EW scale. Unfortunately, the minimal model predicts a tachyonic slepton mass and it must be circumvented in some way, while it is reasonable to think that this property will be lost if other sources of SUSY breaking are introduced simultaneously, since it relies on the fact that the mediator only couples to the Weyl anomaly.

Another well-motivated semi-top-down mechanism is the modulus mediation \cite{Kaplunovsky:1993rd,Brignole:1993dj,Ibanez:1998rf}.
It is a string realization of the gravity mediation based on supergravity \cite{Barbieri:1982eh,Chamseddine:1982jx}.
The modern view of supergravity is the low energy effective theory of superstring. The 10D spacetime is compactified on the Calabi-Yau threefold and the low-energy description of the massless modes is given by the 4D supergravity.
The visible and hidden sectors can be geometrically separated and
 direct interaction between the two sectors could be strongly suppressed \cite{Randall:1998uk,Luty:2001zv,Luty:2001jh,Nelson:2000sn,Nelson:2001mq,Schmaltz:2006qs}\cite{Choi:2006za, Kachru:2007xp}.
On the other hand, the Calabi-Yau manifold is parametrized by the moduli which characterize the size of internal 3 and 4 cycles.
They are 4D massless fields originating from the 6D metric.
Combined with the higher form fields depending on the type of string, they are organized into the chiral multiplets $U_i$, $T_i$ called complex structure moduli and K\"ahler moduli respectively. The 10-dimensional dilaton also produces a massless chiral multiplet $S$ called a complex dilaton. Because of their gravitational origin their interactions with the matter fields are intrinsically nonrenormalizable and often preserve flavor at least at the leading order of $g_s$ and $\alpha^\prime$ expansion \cite{Kaplunovsky:1993rd,Brignole:1993dj,Conlon:2007dw, Choi:2008hn}.
Thus, the modulus and dilaton fields are recognized as good candidates of the SUSY-breaking mediator and their phenomenology has been studied for almost three decades.
In such a study, including more general gravity mediation,
the SUSY spectrum at the EW scale is calculated using the RG equations with the given boundary condition at the string (or Planck) scale. All the relevant interactions between the mediation scale and the EW scale are believed to leave their traces in the low energy parameters via RG running. This machinery sometimes leads to important phenomenological predictions \cite{Inoue:1982pi,Inoue:1983pp,Ibanez:1982fr,Ellis:1982wr,AlvarezGaume:1983gj}\cite{Barbieri:1994pv,Barbieri:1995tw}\cite{Borzumati:1986qx,Hisano:1995nq,Hisano:1995cp,Hisano:1997tc,Hisano:1998fj,Baek:2000sj,Moroi:2000mr,Baek:2001kh}.

In this paper, we argue that the modulus mediation (including dilaton) could have similar properties to the anomaly mediation in some circumstances. The RG running of the soft SUSY breaking parameters due to interactions of massive fields is canceled by their threshold corrections at one-loop order, if their mass is given by nonperturbative dynamics controlled by the same modulus that mediates the SUSY breaking and a sum rule of the modular weights holds for the Yukawa couplings.
Considering intermediate scales in string theory often emerge from such nonperturbative dynamics, this fact could have significant phenomenological implications. As an example, we discuss order reduction of lepton flavor violation in the SUSY seesaw mechanism \cite{Borzumati:1986qx,Hisano:1995nq,Hisano:1995cp,Hisano:1997tc,Hisano:1998fj}.

\section{Modulus mediation}

We first review the modulus mediation of SUSY breaking \footnote{We follow the discussion in \cite{Choi:2005ge, Choi:2005uz}.}.
We employ the superspace notation in Ref.\cite{Wess92} unless otherwise specified.
In the superconformal formulation, the scalar part of $N=1$ 4D supergravity action is given by
\begin{eqnarray}
{\cal S} &=& \int d^4 x  
\sqrt{-g_C}
\left[
\int
d^4\theta \left\{-3 |C|^2 \exp\left(-\frac{\cal K}{3}\right) \right\} 
\right.
\nonumber\\
&+&\left.
\left\{
\int d^2\theta \left(\frac{f_a }{4} W^a W^a +C^3 {\cal W}\right) + {\rm H.c.}
\right\}
+\cdot \cdot \cdot
\right],
\end{eqnarray}
where $C$ denotes the conformal compensator field.
Suppressed terms depend on the curvature or explicit SUSY breaking required for the uplifting of the AdS vacuum as, {\it e.g.}, in the KKLT construction \cite{Kachru:2003aw}. 
In the formula, ${\cal K}$, $f_a$, and ${\cal W}$ represent the K\"ahler potential, gauge kinetic function, and superpotential, respectively.  
The real part of the gauge kinetic function gives the gauge coupling constant:
\begin{equation}
\left.g^2_a = 1/{\rm Re}(f_a)\right|_{\theta^2=\bar{\theta}^2=0}. 
\end{equation}
Fixing the conformal gauge at $C_0 = \exp({\cal K}/6)$ and integrating out $F$-terms using the equation of motion,
\begin{equation}
\label{eq:equationofmotion}
\frac{F^C}{C_0} = \frac{1}{3} F^I {\cal K}_I + m_{3/2},~ F^I = -e^{{\cal K}/2} K^{I\bar{J}} \overline{D_J {\cal W}}.
\end{equation}
we can obtain the component supergravity action in the Einstein frame.
In these formulas the subscript means derivative by the corresponding field and ${\cal K}^{I\bar{J}} = ({\cal K}_{I\bar{J}})^{-1}$, $D_I {\cal W}= {\cal W}_I +{\cal K}_I {\cal W}$ are understood. 
The 4D metric in the conformal frame $g^C_{\mu\nu}$ is related to that of the
 Einstein frame via $g^C_{\mu\nu} = |C|^{-2} \exp({\cal K}/3)\, g^E_{\mu\nu}$.

In the following, we assume that the
K\"ahler potential and the superpotential are expanded
 in terms of the chiral matter superfields $\Phi^i$ in the visible sector as follows:
\begin{eqnarray}
{\cal K} &=& {\cal K}_0(T+T^\dag) + Z_{i}(T+T^\dag) \Phi^i \Phi^{\dag i},\\
{\cal W} &=& W_0 +\frac{1}{2}\mu_{ij}\,\Phi^i\Phi^j
+\frac{1}{3!} \lambda_{ijk}\,\Phi^i \Phi^j \Phi^k, 
\end{eqnarray}
where $T$ denotes the modulus fields collectively.
Here the hidden sector is sequestered from the visible sector. We also explicitly consider the moduli having the shift symmetry $T \to T + i c$ with a real constant $c$ in the Lagrangian, up to total derivatives. 
Thus, the K\"ahler potential has modulus dependence via $T+T^\dag$ and no perturbative $T$ dependence in superpotential due to holomorphy.
For instance, the complex dilaton in the heterotic string and the K\"ahler (complex structure) moduli in type II B (A) orientifolds possess such a property \cite{Kaplunovsky:1993rd,Brignole:1993dj}\cite{Conlon:2007dw}.

The soft SUSY breaking terms of the canonically normalized visible fields can be parametrized as
\begin{eqnarray}
 &&-{\cal L}_{Soft} = m^2_i |\phi|^2 +\left[ \frac{1}{2} M_a \overline{\lambda^a}_L \lambda_L^a\right. \nonumber\\
 &&\left.+\frac{1}{2} B_{ij} M_{ij} \phi^i \phi^j 
+ \frac{1}{3!} A_{ijk} Y_{ijk} \phi^i \phi^j \phi^k
+{\rm H.c.} \right],
\end{eqnarray}
where canonical mass parameters and the Yukawa couplings are given by
\begin{equation}
M_{ij} = \mu_{ij}/\sqrt{ Y_i Y_j},~
Y_{ijk} = \lambda_{ijk}/\sqrt{Y_i Y_j Y_k}, 
\end{equation}
with $Y_i = e^{-{\cal K}_0/3} Z_i$.
These terms can be read off by integrating out the $F$-terms of the visible fields $\Phi^i$ in the supergravity action using their equation of motion as
\begin{eqnarray}
\label{eq:softmass1}
&& M_a = F^I \partial_I \ln \left({\rm Re} f_a\right),~~
m_i^2 = - F^I F^{\bar{J}} \partial_I \partial_{\bar{J}} \ln Y_i,
\\
&& A_{ijk} = A_i + A_j + A_k-F^I\partial_I \ln\left(\lambda_{ijk}\right), \\
\label{eq:softmass4}
&&B_{ij} = -\frac{F^C}{C_0}+A_i + A_j-F^I\partial_I \ln\left(\mu_{ij}\right),  
\end{eqnarray}
where $A_i = F^I\partial_I \ln Y_i$.
$F^I$, and $\partial_I$ denote the $F$-terms and the derivative for the moduli (or hidden sector fields in general).
In the formulas, we leave the field dependence in the holomorphic couplings for later purposes.
We also omit the effects of the flavor mixing to avoid unnecessary complications. Their inclusion by the K\"ahler connection is straightforward and not essential in the following discussion.
We will consider the gauge modulus having the following tree-level property at the string scale,
\begin{equation}
\label{eq:string-bc} 
f_a = T,~~~~~Y_i = (T+T^\dag)^{c_i}.
\end{equation}
The complex dilaton in heterotic string and the overall K\"ahler modulus in type II B toroidal orientifolds are among the examples.
$c_i$ is related to the modular weight of the matter field $n_i$ as $c_i=1-n_i$ \cite{Choi:2005ge,Choi:2005uz} and typically given by a ratio of small integers depending on the origin of the field.
Putting them into Eqs. (\ref{eq:softmass1})--(\ref{eq:softmass4}), we obtain 
\begin{equation}
 \label{eq:GUTBC}
 M_a = M_0,~~ A_{ijk} = (c_i+c_j+c_k) M_0, ~~m_i^2 = c_i |M_0|^2,  
\end{equation}
with $M_0 \equiv F^T/(T+T^\dag)$ at the string scale.
For the phenomenological purpose, we need the values evolved by the renormalization group equations down to the EW scale.
There is also contribution from the anomaly mediation radiatively mediated by $C$.
We suppress the anomaly mediation as a subleading contribution; however, it is straightforward to check that the following discussions also held under the simultaneous presence of anomaly mediation (the mirage mediation) \cite{Choi:2004sx,Choi:2005ge,Choi:2005uz,Choi:2005hd}\cite{Endo:2005uy}.

\section{Method of similarity}

We discuss evolution of the soft SUSY breaking parameters under the 1-loop RG equations in terms of the method of similarity \cite{Arnold78, Landau76, Kitano:2006gv}.
The 1-loop RG equations for gauge coupling and the matter K\"ahler metric is given by
\begin{eqnarray}
\label{eq:betafunction}
\frac{d\, g_a^{-2}}{d \ln \mu}  &=& -\frac{b_a}{8\pi^2},~~  b_a = -3\,
T_a^G+\sum_i T_a^{i}, \\
\label{eq:anomalousdimension}
\frac{d\, \ln Y_i }{d \ln \mu}  &=& \frac{\gamma_i}{8\pi^2},~
\gamma_i = 2 \sum_a C_2^{a\,i} g_a^2 - \frac{1}{2} \sum_{jk} |Y_{ijk}|^2,
\end{eqnarray}
where $T_a^G$ ($T_a^{i}$) is the Dynkin index of the adjoint ($\phi^i$) representation for the corresponding gauge group.
$C_2^{a\, i}$ denotes the quadratic Casimir of $\phi^i$. 
This results in the RG equation for the canonical Yukawa coupling,
\begin{equation}
 \frac{d\, Y_{ijk}}{d \ln \mu}  = -\frac{1}{16\pi^2}\left(\gamma_i + \gamma_j + \gamma_k \right) Y_{ijk},
\end{equation}
 since the holomorphic couplings are not renormalized.

It is known that these first order differential equations are invariant under the following scaling transformation with a real constant $\Delta$,  
\begin{equation}
\ln \mu \to \Delta\multi \ln \mu,~~g_a^{-2} \to \Delta\multi g_a^{-2},~~Y_{ijk} \to \Delta^{-\frac{1}{2}}\multi Y_{ijk}, 
\end{equation}
Then, if $g_a^{-2}[\,g_{a\,\String}^{-2};\, (\mu/M_\String)\,]$ represents a solution for the boundary condition, $g_a^{-2} = g_{a\,\String}^{-2}$ at $\mu = M_\String$, a function, $\Delta^{-1}\multi g_a^{-2}[\,\Delta\multi g_{a\,\String}^{-2};\, (\mu/M_\String)^\Delta\,]$ also satisfies the equation. The boundary condition of this solution is $g_{a\,\String}^{-2}$, therefore, the uniqueness of the solution leads to
\begin{equation}
 g_a^{-2}[\, g_{a\,\String}^{-2};\,(\mu/M_\String)\,] = \Delta^{-1}\multi g_a^{-2}[\,\Delta\multi g_{a\,\String}^{-2};\,(\mu/M_\String)^\Delta\,].
\end{equation}
A similar discussion for the Yukawa coupling follows:
\begin{eqnarray}
&& Y_{ijk}[\, \{ g_{a\,\String},\, Y_\String \};\, (\mu/M_\String)\,] = \nonumber\\
&& ~~~~~\Delta^{\frac{1}{2}}\multi Y_{ijk}[\, \{\Delta^{-\frac{1}{2}}\multi g_{a\,\String},\, \Delta^{-\frac{1}{2}}\multi Y_\String \};\, (\mu/M_\String)^\Delta\,],
\end{eqnarray}
where $\{g_{a\,\String},\, Y_\String\}$ represents the boundary condition of the relevant gauge and Yukawa couplings for the set of RG equations. 
In the following, we take $M_\String$ as the string scale where the boundary condition, Eq. (\ref{eq:string-bc}) is satisfied.
If the sum rule, $a_{ijk} = c_i + c_j + c_k =1$ holds for all the Yukawa couplings, which is often the case for the large Yukawa couplings \cite{Burgess:1985zz,Nilles:1986cy,Choi:1987is,Conlon:2006tj,Choi:2006za,Choi:2008hn}, 
 we set $\Delta$ as
\begin{equation}
 \Delta = \langle T+T^\dag\rangle/(T+T^\dag),
\end{equation}
where the rectangle parentheses denote the vacuum expectation value.
Then we obtain the final expressions,
\begin{equation}
 g_a^{-2}[\, g_{a\,\String}^{-2};\,(\mu/M_\String)\,] = \Delta^{-1}\multi g_a^{-2}[\, \langle g_{a\,\String}^{-2} \rangle ; \,(\mu/M_\String)^\Delta\,],
\end{equation}  
\begin{eqnarray}
&& Y_{ijk}[\, \{g_{a\,\String},\, Y_\String \};\, (\mu/M_\String)\,] = \nonumber\\
&& ~~~~~\Delta^{\frac{1}{2}}\multi Y_{ijk}[\, \{\langle g_{a\,\String} \rangle,\  \langle Y_\String \rangle \}  ;\, (\mu/M_\String)^\Delta\,].
\end{eqnarray}
In these formulas, the modulus dependence originally residing in the boundary conditions are squeezed into the overall factors and the renormalization scale dependence.
Then the analytic continuation into superspace \cite{Giudice:1997ni, ArkaniHamed:1998kj} and Eqs. (\ref{eq:softmass1})--(\ref{eq:softmass4}) with $Z_i$ obtained by formal integration of $\gamma_i$ yields a closed form for the soft SUSY breaking parameters at the scale $\mu$ using the SUSY parameters at the same scale \cite{Choi:2005uz}, 
\begin{eqnarray}
\label{eq:mirage1}
&&\frac{M_a(\mu)}{M_0} = 1 + \frac{b_a}{8\pi^2}g_a^2(\mu)\, L,\\
&&\frac{A_{ijk}(\mu)}{M_0} = a_{ijk}-\frac{1}{8\pi^2}\left(\gamma_i+\gamma_j+\gamma_k\right)\, L,\\
\label{eq:mirage3}
%
%
&&\frac{{m_i^2}(\mu)}{|M_0|^2} = c_i 
-\frac{1}{4\pi^2}\gamma_i\, L-\frac{1}{8\pi^2}\frac{d \gamma_i}{d \ln \mu}\, L^2, 
\end{eqnarray}
where $L=\ln(\mu/M_\String)$.
This property is analogous to the anomaly mediation where $C$ dependence is encapsulated in the renormalization scale dependence as, $\mu/(CM_\String)$ after the redefinition of the visible fields, $C \Phi_i \to \Phi_i$, however, only valid within the one-loop approximation.

\section{Trace of massive fields}

In this paper, we will examine the corrections to the soft SUSY breaking parameters due to massive fields integrated out at some scale. As an example, we consider chiral superfields $\Psi^\alpha$ $(\alpha=1,2)$ having a mass $X$ and the Yukawa couplings $\lambda^\prime$ in the superpotential as
\begin{equation}
{\delta {\cal W}} = X \Psi^1 \Psi^2 + \frac{1}{2} \lambda^\prime_{ij\alpha} \Phi^i \Phi^j \Psi^\alpha + \frac{1}{2} \lambda^\prime_{i\alpha\beta} \Phi^i \Psi^\alpha \Psi^\beta,
\end{equation}
where $\Phi^i$ represents fields remain in the low energy effective theory.
At the threshold $\mu=X$, we connect the RG solutions for their gauge and Yukawa couplings continuously, while switching the gauge beta function using a step function \footnote{Precisely, the threshold is given by $X/\sqrt{Z_{\Psi^1}\,Z_{\Psi^2}}$. The difference in the estimate of the SUSY breaking appears to be the next-to-leading logarithm order since picking up a $F$ component in $Z_{\Psi^\alpha}$ eliminates one logarithm.}.
Applying the formula in the previous section recursively, we obtain,
\begin{eqnarray}
 && g_a^{-2} [\,g_{a\,X}^{-2};\,(\mu/|X|)\,] \nonumber\\
 &&=\Delta^{-1}\multi g_a^{-2} [\,g_{a\,X}^{-2}[\,\Delta\multi g_{a\, \String}^{-2};\,(|X|/M_\String)^\Delta];\,(\mu/|X|)^\Delta\,] \nonumber\\
 &&= \Delta^{-1}\multi g_a^{-2} [\,\Delta\multi g_{a\, \String}^{-2};\,(|X|/M_\String)^\Delta,\, (\mu/M_\String)^\Delta\,]\nonumber\\
 &&= \Delta^{-1}\multi g_a^{-2} [\,\langle g_{a\, \String}^{-2} \rangle;\,(|X|/M_\String)^\Delta,\, (\mu/M_\String)^\Delta\,],
\end{eqnarray}
where the subscript $X$ stands for the value at $\mu=X$.
A similar expression holds for the Yukawa coupling,
\begin{eqnarray}  
&& Y_{ijk}[\,\{g_{a\,X},\,Y_X\};\,(\mu/|X|)\,] \nonumber\\
 && =\Delta^{\frac{1}{2}}\multi Y_{ijk}[\,\{\langle g_{a\,\String} \rangle, \Delta^{(a-1)/2}\multi \langle Y_\String \rangle ,\Delta^{(a^\prime-1)/2}\multi \langle Y_\String^\prime \rangle \}; \nonumber\\
&&\phantom{=\Delta^{\frac{1}{2}} Y_{ijk}[\,\{\langle g_{a\,\String} \rangle,}
 \,(|X|/M_\String)^\Delta, \,(\mu/M_\String)^\Delta\,],
\end{eqnarray}
where $Y^\prime$ represents the canonical Yukawa coupling corresponds to $\lambda^\prime$.

If the sum rule of $c_i$ holds for all the Yukawa couplings ($a_{ijk} = a^\prime_{ij\alpha}=a^\prime_{i\alpha\beta}=1$), the modulus dependence can be squeezed into the overall factor and the dependence on the threshold mass and the renormalization scale. Furthermore, if the threshold is given by the nonperturbative effect like
\begin{equation}
 \label{eq:non-perturbative-mass}
 X = \Lambda \exp\left(-b T\right),
\end{equation}
and $\Lambda$ is given by the cutoff $\Lambda = M_\String$, the threshold dependence disappears.
Note that such a nonperturbative mass due to field theoretic or string instantons is a typical origin of the intermediate scale in string theory \cite{Kim:1991mv,Chun:1991xm,Faraggi:1993zh,Coriano:2003ui}\cite{Blumenhagen:2006xt,Ibanez:2007rs,Florea:2006si} (see also Ref. \cite{Langacker:2011bi} and the references therein). 
In such a case, the low energy soft SUSY breaking parameters again given by the closed form at $\mu$ as in Eqs. (\ref{eq:mirage1})--(\ref{eq:mirage3}). This kind of ultraviolet insensitivity means that the threshold correction integrating out $\Psi^\alpha$ cancels the effect of RG running due to $Y^\prime$ as in the anomaly mediation \footnote{See also \cite{Pomarol:1999ie,Rattazzi:1999qg,Okada:2002mv,Everett:2008qy,Everett:2008ey,Choi:2009jn,Altunkaynak:2010xe} for studies on the threshold corrections in the modulus and anomaly mediation.}.
The leftover is suppressed by $\ln(\Lambda/M_\String)/8\pi^2$.
If some Yukawa coupling $Y$ does not satisfies the sum rule, the leading effect of $Y^\prime$ in the soft SUSY breaking appears through the interference terms like, $|Y_\String|^2 |Y_\String^\prime|^2 [\ln(|\langle X \rangle|/M_\String)/8\pi^2]^2$ in the K\"ahler metric, which is loop suppressed while $\log(|\langle X \rangle|/M_\String)$ enhanced.
Inclusion of multithresholds is straightforward and we will not repeat the discussion here.
The massive field can be a gauge multiplet spontaneously broken by the scalar with a nonperturbative mass.

\section{Example}

As a phenomenological example, we discuss lepton flavor violation in the SUSY seesaw mechanism \cite{Borzumati:1986qx,Hisano:1995nq,Hisano:1995cp,Hisano:1997tc,Hisano:1998fj}.
We introduce three generations of right-hand neutrinos $\bar{N}_i$ in the MSSM with the superpotential
\begin{equation}
\delta W = \frac{1}{2} M_R \bar{N}_i \bar{N}_i + \lambda_{\nu}^{ij} H_u L_i \bar{N}_j, 
\end{equation}
where $H_u$, $L_i$ are up-type Higgs and lepton doublets in the MSSM (MSSMRN).
We choose a flavor universal Majorana mass for simplicity.
After integrating out $\bar{N}_i$, small Majorana neutrino masses are generated through dimension 5 operators \cite{Yanagida:1979as,GellMann:1980vs}.
In this scenario, it is well known that the RG running due to the neutrino Yukawa coupling induces flavor mixing in the left-hand slepton mass matrix, which results in the lepton flavor violating processes like $\mu \to e \gamma$ forbidden in the SM.

\begin{figure}[t]
\begin{center}
 \includegraphics[clip,width=8.6cm]{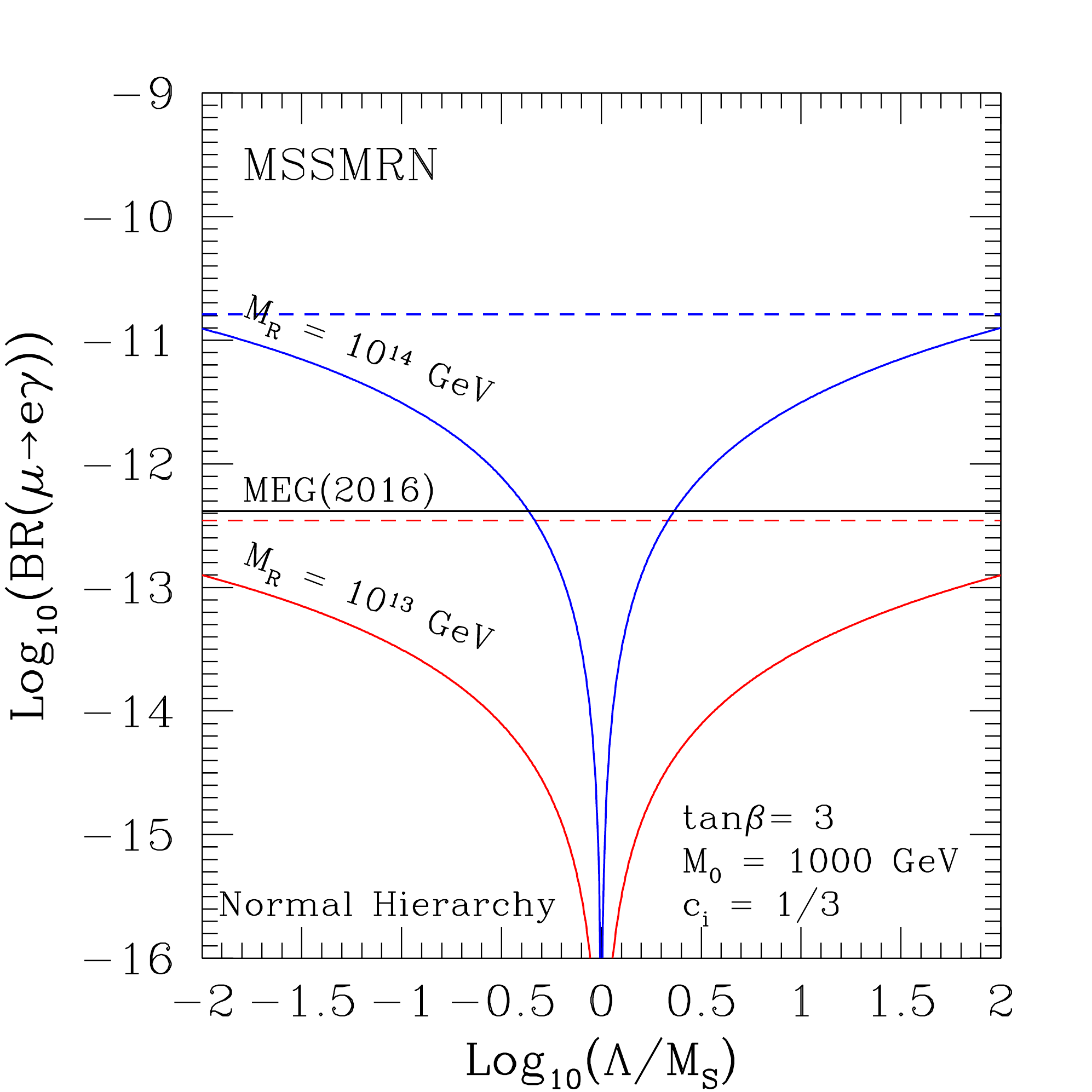}%
  \caption{\label{fig:mueg} $BR(\mu \to e \gamma)$ in the MSSM with right-handed neutrinos. The solid (dashed) curve represents the result for the nonperturbative (constant) mass term. The gluino mass is 2.3 TeV at the one-loop level. }
\end{center}   
\end{figure}

\begin{figure}[t]
\begin{center}
  \includegraphics[clip,width=8.6cm]{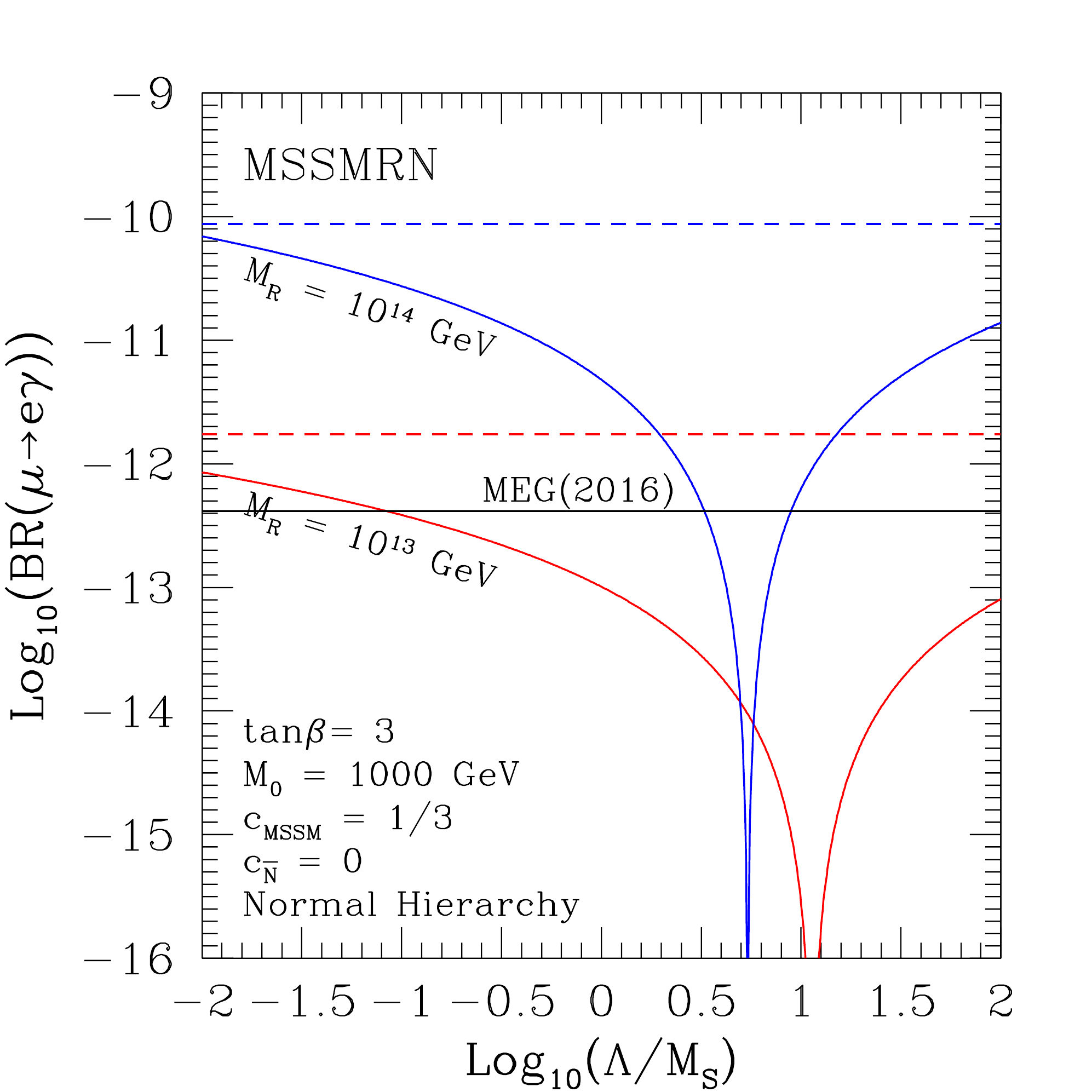}%
  \caption{\label{fig:mueg2} The same as in Fig. 1 except the sum rule is violated in the neutrino Yukawa coupling ($c_{\bar{N}=0}$).}
\end{center}
\end{figure}

In Fig.\ref{fig:mueg}, we calculate $BR(\mu \to e \gamma)$ for the constant $M_R$ (dotted line) and $M_R$ given by Eq. (\ref{eq:non-perturbative-mass}) (solid curve) as a function of $\log_{10}(\Lambda/M_\String)$.  
We set the boundary condition, Eq. (\ref{eq:GUTBC}) at the unification scale, $M_G=2\times 10^{16}$ GeV identified with the string scale and evolve the soft parameters down to the EW scale.
We choose the modular weights of the MSSM fields as $c_i=1/3$ so that all the Yukawa couplings satisfy the sum rule.
The universal gaugino mass is set to 1 TeV, which corresponds to $2.3$ TeV gluino with the one-loop RG equation. 
The threshold correction at $M_R$ is calculated following the method in Refs. \cite{Giudice:1997ni, ArkaniHamed:1998kj}.
For the estimation of the neutrino mass matrix, we adopt the normal hierarchy and the result of the global fit in Ref. \cite{Esteban:2018azc} with the assumption of vanishing lightest neutrino mass.
The horizontal line in the figure indicates the current experimental bound \cite{TheMEG:2016wtm}.
We show two cases, $M_R$=$10^{13},\, 10^{14}$ GeV. In both cases, the branching ratio for the nonperturbative threshold disappears at $\Lambda=M_\String$ due to cancellation. Even an order difference in $\Lambda/M_\String$ can lead to an order reduction in $BR(\mu \to e \gamma)$.
This revives the parameter space already excluded by the experimental bound.
In the calculation of $BR(\mu \to e \gamma)$, we confirm that the sum rule for the top Yukawa coupling is numerically irrelevant, while breaking it in the neutrino Yukawa coupling shifts the point of cancellation from $\Lambda=M_\String$ as shown in Fig.\ref{fig:mueg2}, where we take $c_{\bar{N}}=0$ with the other modular weights and mass parameters are intact.

\section{Conclusion}

We argued that the RG running of the soft SUSY breaking parameters due to interaction of massive fields is canceled by their threshold correction at one-loop order in the modulus mediation if their mass is given by nonperturbative dynamics controlled by the same modulus that mediates SUSY breaking and the modular-weight sum rule is satisfied in the case of the Yukawa couplings. This could have significant implications for phenomenology. As an example, we showed that lepton flavor violation in the SUSY seesaw mechanism could have order reduction.
The cancellation also works with the anomaly mediation and it is interesting to introduce the seesaw mechanism or vectorlike fields in the TeV scale mirage mediation models \cite{Choi:2005hd}\cite{Kitano:2005wc,Kitano:2006gv,Choi:2006xb,Abe:2007je,Kobayashi:2012ee,Asano:2012sv,Hagimoto:2015tua,Kawamura:2017qey} without spoiling its little SUSY hierarchy.
There might be a plethora of other phenomenological applications including massive gauge bosons as in the grand unified theories.
We also note that the mechanism itself is not limited to low energy SUSY models. The scale of the soft SUSY breaking is arbitrary.
On the other hand, exploration of string model space realizing the cancellation is also an interesting issue, although it is beyond the scope of this paper. We leave them to future work.
 
We acknowledge helpful discussion with Kiwoon Choi.
The author is also grateful to Kazuto Uenou for his contribution in the early stages of this study.
The numerical analysis is partly performed with XC40 at YITP.
K.O. is supported in part by RCAPP and RCSHE at Kyushu University, MEXT Japan and Public Interest Incorporated Foundation Fuujyukai. 


%

%



\begin{acknowledgments}
\end{acknowledgments}

\bibliography{KiO1}

\providecommand{\noopsort}[1]{}\providecommand{\singleletter}[1]{#1}%
\begin{thebibliography}{79}%
\makeatletter
\providecommand \@ifxundefined [1]{%
 \@ifx{#1\undefined}
}%
\providecommand \@ifnum [1]{%
 \ifnum #1\expandafter \@firstoftwo
 \else \expandafter \@secondoftwo
 \fi
}%
\providecommand \@ifx [1]{%
 \ifx #1\expandafter \@firstoftwo
 \else \expandafter \@secondoftwo
 \fi
}%
\providecommand \natexlab [1]{#1}%
\providecommand \enquote  [1]{``#1''}%
\providecommand \bibnamefont  [1]{#1}%
\providecommand \bibfnamefont [1]{#1}%
\providecommand \citenamefont [1]{#1}%
\providecommand \href@noop [0]{\@secondoftwo}%
\providecommand \href [0]{\begingroup \@sanitize@url \@href}%
\providecommand \@href[1]{\@@startlink{#1}\@@href}%
\providecommand \@@href[1]{\endgroup#1\@@endlink}%
\providecommand \@sanitize@url [0]{\catcode `\\12\catcode `\$12\catcode
  `\&12\catcode `\#12\catcode `\^12\catcode `\_12\catcode `\%12\relax}%
\providecommand \@@startlink[1]{}%
\providecommand \@@endlink[0]{}%
\providecommand \url  [0]{\begingroup\@sanitize@url \@url }%
\providecommand \@url [1]{\endgroup\@href {#1}{\urlprefix }}%
\providecommand \urlprefix  [0]{URL }%
\providecommand \Eprint [0]{\href }%
\providecommand \doibase [0]{https://doi.org/}%
\providecommand \selectlanguage [0]{\@gobble}%
\providecommand \bibinfo  [0]{\@secondoftwo}%
\providecommand \bibfield  [0]{\@secondoftwo}%
\providecommand \translation [1]{[#1]}%
\providecommand \BibitemOpen [0]{}%
\providecommand \bibitemStop [0]{}%
\providecommand \bibitemNoStop [0]{.\EOS\space}%
\providecommand \EOS [0]{\spacefactor3000\relax}%
\providecommand \BibitemShut  [1]{\csname bibitem#1\endcsname}%
\let\auto@bib@innerbib\@empty
\bibitem [{\citenamefont {Nilles}(1984)}]{Nilles:1983ge}%
  \BibitemOpen
  \bibfield  {author} {\bibinfo {author} {\bibfnamefont {H.~P.}\ \bibnamefont
  {Nilles}},\ }\bibfield  {title} {\bibinfo {title} {{Supersymmetry,
  Supergravity and Particle Physics}},\ }\href
  {https://doi.org/10.1016/0370-1573(84)90008-5} {\bibfield  {journal}
  {\bibinfo  {journal} {Phys. Rept.}\ }\textbf {\bibinfo {volume} {110}},\
  \bibinfo {pages} {1} (\bibinfo {year} {1984})}\BibitemShut {NoStop}%
\bibitem [{\citenamefont {Martin}(1997)}]{Martin:1997ns}%
  \BibitemOpen
  \bibfield  {author} {\bibinfo {author} {\bibfnamefont {S.~P.}\ \bibnamefont
  {Martin}},\ }\bibfield  {title} {\bibinfo {title} {{A Supersymmetry
  primer}},\ }\href {https://doi.org/10.1142/9789812839657_0001,
  10.1142/9789814307505_0001} {\ ,\ \bibinfo {pages} {1} (\bibinfo {year}
  {1997})},\ \bibinfo {note} {[Adv. Ser. Direct. High Energy
  Phys.18,1(1998)]},\ \Eprint {https://arxiv.org/abs/hep-ph/9709356}
  {arXiv:hep-ph/9709356 [hep-ph]} \BibitemShut {NoStop}%
\bibitem [{\citenamefont {Giudice}\ and\ \citenamefont
  {Rattazzi}(1999)}]{Giudice:1998bp}%
  \BibitemOpen
  \bibfield  {author} {\bibinfo {author} {\bibfnamefont {G.~F.}\ \bibnamefont
  {Giudice}}\ and\ \bibinfo {author} {\bibfnamefont {R.}~\bibnamefont
  {Rattazzi}},\ }\bibfield  {title} {\bibinfo {title} {{Theories with gauge
  mediated supersymmetry breaking}},\ }\href
  {https://doi.org/10.1016/S0370-1573(99)00042-3} {\bibfield  {journal}
  {\bibinfo  {journal} {Phys. Rept.}\ }\textbf {\bibinfo {volume} {322}},\
  \bibinfo {pages} {419} (\bibinfo {year} {1999})},\ \Eprint
  {https://arxiv.org/abs/hep-ph/9801271} {arXiv:hep-ph/9801271 [hep-ph]}
  \BibitemShut {NoStop}%
\bibitem [{\citenamefont {Randall}\ and\ \citenamefont
  {Sundrum}(1999)}]{Randall:1998uk}%
  \BibitemOpen
  \bibfield  {author} {\bibinfo {author} {\bibfnamefont {L.}~\bibnamefont
  {Randall}}\ and\ \bibinfo {author} {\bibfnamefont {R.}~\bibnamefont
  {Sundrum}},\ }\bibfield  {title} {\bibinfo {title} {{Out of this world
  supersymmetry breaking}},\ }\href
  {https://doi.org/10.1016/S0550-3213(99)00359-4} {\bibfield  {journal}
  {\bibinfo  {journal} {Nucl. Phys.}\ }\textbf {\bibinfo {volume} {B557}},\
  \bibinfo {pages} {79} (\bibinfo {year} {1999})},\ \Eprint
  {https://arxiv.org/abs/hep-th/9810155} {arXiv:hep-th/9810155 [hep-th]}
  \BibitemShut {NoStop}%
\bibitem [{\citenamefont {Giudice}\ \emph {et~al.}(1998)\citenamefont
  {Giudice}, \citenamefont {Luty},\ and\ \citenamefont
  {Murayama}}]{Giudice:1998xp}%
  \BibitemOpen
  \bibfield  {author} {\bibinfo {author} {\bibfnamefont {G.~F.}\ \bibnamefont
  {Giudice}}, \bibinfo {author} {\bibfnamefont {M.~A.}\ \bibnamefont {Luty}},\
  and\ \bibinfo {author} {\bibfnamefont {H.}~\bibnamefont {Murayama}},\
  }\bibfield  {title} {\bibinfo {title} {{Gaugino mass without singlets}},\
  }\href {https://doi.org/10.1088/1126-6708/1998/12/027} {\bibfield  {journal}
  {\bibinfo  {journal} {JHEP}\ }\textbf {\bibinfo {volume} {12}},\ \bibinfo
  {pages} {027}},\ \Eprint {https://arxiv.org/abs/hep-ph/9810442}
  {arXiv:hep-ph/9810442 [hep-ph]} \BibitemShut {NoStop}%
\bibitem [{\citenamefont {Giudice}\ and\ \citenamefont
  {Rattazzi}(1998)}]{Giudice:1997ni}%
  \BibitemOpen
  \bibfield  {author} {\bibinfo {author} {\bibfnamefont {G.~F.}\ \bibnamefont
  {Giudice}}\ and\ \bibinfo {author} {\bibfnamefont {R.}~\bibnamefont
  {Rattazzi}},\ }\bibfield  {title} {\bibinfo {title} {{Extracting
  supersymmetry breaking effects from wave function renormalization}},\ }\href
  {https://doi.org/10.1016/S0550-3213(97)00647-0} {\bibfield  {journal}
  {\bibinfo  {journal} {Nucl. Phys.}\ }\textbf {\bibinfo {volume} {B511}},\
  \bibinfo {pages} {25} (\bibinfo {year} {1998})},\ \Eprint
  {https://arxiv.org/abs/hep-ph/9706540} {arXiv:hep-ph/9706540 [hep-ph]}
  \BibitemShut {NoStop}%
\bibitem [{\citenamefont {Arkani-Hamed}\ \emph {et~al.}(1998)\citenamefont
  {Arkani-Hamed}, \citenamefont {Giudice}, \citenamefont {Luty},\ and\
  \citenamefont {Rattazzi}}]{ArkaniHamed:1998kj}%
  \BibitemOpen
  \bibfield  {author} {\bibinfo {author} {\bibfnamefont {N.}~\bibnamefont
  {Arkani-Hamed}}, \bibinfo {author} {\bibfnamefont {G.~F.}\ \bibnamefont
  {Giudice}}, \bibinfo {author} {\bibfnamefont {M.~A.}\ \bibnamefont {Luty}},\
  and\ \bibinfo {author} {\bibfnamefont {R.}~\bibnamefont {Rattazzi}},\
  }\bibfield  {title} {\bibinfo {title} {{Supersymmetry breaking loops from
  analytic continuation into superspace}},\ }\href
  {https://doi.org/10.1103/PhysRevD.58.115005} {\bibfield  {journal} {\bibinfo
  {journal} {Phys. Rev.}\ }\textbf {\bibinfo {volume} {D58}},\ \bibinfo {pages}
  {115005} (\bibinfo {year} {1998})},\ \Eprint
  {https://arxiv.org/abs/hep-ph/9803290} {arXiv:hep-ph/9803290 [hep-ph]}
  \BibitemShut {NoStop}%
\bibitem [{\citenamefont {Kaplunovsky}\ and\ \citenamefont
  {Louis}(1993)}]{Kaplunovsky:1993rd}%
  \BibitemOpen
  \bibfield  {author} {\bibinfo {author} {\bibfnamefont {V.~S.}\ \bibnamefont
  {Kaplunovsky}}\ and\ \bibinfo {author} {\bibfnamefont {J.}~\bibnamefont
  {Louis}},\ }\bibfield  {title} {\bibinfo {title} {{Model independent analysis
  of soft terms in effective supergravity and in string theory}},\ }\href
  {https://doi.org/10.1016/0370-2693(93)90078-V} {\bibfield  {journal}
  {\bibinfo  {journal} {Phys. Lett.}\ }\textbf {\bibinfo {volume} {B306}},\
  \bibinfo {pages} {269} (\bibinfo {year} {1993})},\ \Eprint
  {https://arxiv.org/abs/hep-th/9303040} {arXiv:hep-th/9303040 [hep-th]}
  \BibitemShut {NoStop}%
\bibitem [{\citenamefont {Brignole}\ \emph {et~al.}(1994)\citenamefont
  {Brignole}, \citenamefont {Ibanez},\ and\ \citenamefont
  {Munoz}}]{Brignole:1993dj}%
  \BibitemOpen
  \bibfield  {author} {\bibinfo {author} {\bibfnamefont {A.}~\bibnamefont
  {Brignole}}, \bibinfo {author} {\bibfnamefont {L.~E.}\ \bibnamefont
  {Ibanez}},\ and\ \bibinfo {author} {\bibfnamefont {C.}~\bibnamefont
  {Munoz}},\ }\bibfield  {title} {\bibinfo {title} {{Towards a theory of soft
  terms for the supersymmetric Standard Model}},\ }\href
  {https://doi.org/10.1016/0550-3213(94)00600-J, 10.1016/0550-3213(94)00068-9}
  {\bibfield  {journal} {\bibinfo  {journal} {Nucl. Phys.}\ }\textbf {\bibinfo
  {volume} {B422}},\ \bibinfo {pages} {125} (\bibinfo {year} {1994})},\
  \bibinfo {note} {[Erratum: Nucl. Phys.B436,747(1995)]},\ \Eprint
  {https://arxiv.org/abs/hep-ph/9308271} {arXiv:hep-ph/9308271 [hep-ph]}
  \BibitemShut {NoStop}%
\bibitem [{\citenamefont {Ibanez}\ \emph {et~al.}(1999)\citenamefont {Ibanez},
  \citenamefont {Munoz},\ and\ \citenamefont {Rigolin}}]{Ibanez:1998rf}%
  \BibitemOpen
  \bibfield  {author} {\bibinfo {author} {\bibfnamefont {L.~E.}\ \bibnamefont
  {Ibanez}}, \bibinfo {author} {\bibfnamefont {C.}~\bibnamefont {Munoz}},\ and\
  \bibinfo {author} {\bibfnamefont {S.}~\bibnamefont {Rigolin}},\ }\bibfield
  {title} {\bibinfo {title} {{Aspect of type I string phenomenology}},\ }\href
  {https://doi.org/10.1016/S0550-3213(99)00264-3} {\bibfield  {journal}
  {\bibinfo  {journal} {Nucl. Phys.}\ }\textbf {\bibinfo {volume} {B553}},\
  \bibinfo {pages} {43} (\bibinfo {year} {1999})},\ \Eprint
  {https://arxiv.org/abs/hep-ph/9812397} {arXiv:hep-ph/9812397 [hep-ph]}
  \BibitemShut {NoStop}%
\bibitem [{\citenamefont {Barbieri}\ \emph {et~al.}(1982)\citenamefont
  {Barbieri}, \citenamefont {Ferrara},\ and\ \citenamefont
  {Savoy}}]{Barbieri:1982eh}%
  \BibitemOpen
  \bibfield  {author} {\bibinfo {author} {\bibfnamefont {R.}~\bibnamefont
  {Barbieri}}, \bibinfo {author} {\bibfnamefont {S.}~\bibnamefont {Ferrara}},\
  and\ \bibinfo {author} {\bibfnamefont {C.~A.}\ \bibnamefont {Savoy}},\
  }\bibfield  {title} {\bibinfo {title} {{Gauge Models with Spontaneously
  Broken Local Supersymmetry}},\ }\href
  {https://doi.org/10.1016/0370-2693(82)90685-2} {\bibfield  {journal}
  {\bibinfo  {journal} {Phys. Lett.}\ }\textbf {\bibinfo {volume} {119B}},\
  \bibinfo {pages} {343} (\bibinfo {year} {1982})}\BibitemShut {NoStop}%
\bibitem [{\citenamefont {Chamseddine}\ \emph {et~al.}(1982)\citenamefont
  {Chamseddine}, \citenamefont {Arnowitt},\ and\ \citenamefont
  {Nath}}]{Chamseddine:1982jx}%
  \BibitemOpen
  \bibfield  {author} {\bibinfo {author} {\bibfnamefont {A.~H.}\ \bibnamefont
  {Chamseddine}}, \bibinfo {author} {\bibfnamefont {R.~L.}\ \bibnamefont
  {Arnowitt}},\ and\ \bibinfo {author} {\bibfnamefont {P.}~\bibnamefont
  {Nath}},\ }\bibfield  {title} {\bibinfo {title} {{Locally Supersymmetric
  Grand Unification}},\ }\href {https://doi.org/10.1103/PhysRevLett.49.970}
  {\bibfield  {journal} {\bibinfo  {journal} {Phys. Rev. Lett.}\ }\textbf
  {\bibinfo {volume} {49}},\ \bibinfo {pages} {970} (\bibinfo {year}
  {1982})}\BibitemShut {NoStop}%
\bibitem [{\citenamefont {Luty}\ and\ \citenamefont
  {Sundrum}(2003)}]{Luty:2001zv}%
  \BibitemOpen
  \bibfield  {author} {\bibinfo {author} {\bibfnamefont {M.}~\bibnamefont
  {Luty}}\ and\ \bibinfo {author} {\bibfnamefont {R.}~\bibnamefont {Sundrum}},\
  }\bibfield  {title} {\bibinfo {title} {{Anomaly mediated supersymmetry
  breaking in four-dimensions, naturally}},\ }\href
  {https://doi.org/10.1103/PhysRevD.67.045007} {\bibfield  {journal} {\bibinfo
  {journal} {Phys. Rev.}\ }\textbf {\bibinfo {volume} {D67}},\ \bibinfo {pages}
  {045007} (\bibinfo {year} {2003})},\ \Eprint
  {https://arxiv.org/abs/hep-th/0111231} {arXiv:hep-th/0111231 [hep-th]}
  \BibitemShut {NoStop}%
\bibitem [{\citenamefont {Luty}\ and\ \citenamefont
  {Sundrum}(2002)}]{Luty:2001jh}%
  \BibitemOpen
  \bibfield  {author} {\bibinfo {author} {\bibfnamefont {M.~A.}\ \bibnamefont
  {Luty}}\ and\ \bibinfo {author} {\bibfnamefont {R.}~\bibnamefont {Sundrum}},\
  }\bibfield  {title} {\bibinfo {title} {{Supersymmetry breaking and composite
  extra dimensions}},\ }\href {https://doi.org/10.1103/PhysRevD.65.066004}
  {\bibfield  {journal} {\bibinfo  {journal} {Phys. Rev.}\ }\textbf {\bibinfo
  {volume} {D65}},\ \bibinfo {pages} {066004} (\bibinfo {year} {2002})},\
  \Eprint {https://arxiv.org/abs/hep-th/0105137} {arXiv:hep-th/0105137
  [hep-th]} \BibitemShut {NoStop}%
\bibitem [{\citenamefont {Nelson}\ and\ \citenamefont
  {Strassler}(2000)}]{Nelson:2000sn}%
  \BibitemOpen
  \bibfield  {author} {\bibinfo {author} {\bibfnamefont {A.~E.}\ \bibnamefont
  {Nelson}}\ and\ \bibinfo {author} {\bibfnamefont {M.~J.}\ \bibnamefont
  {Strassler}},\ }\bibfield  {title} {\bibinfo {title} {{Suppressing flavor
  anarchy}},\ }\href {https://doi.org/10.1088/1126-6708/2000/09/030} {\bibfield
   {journal} {\bibinfo  {journal} {JHEP}\ }\textbf {\bibinfo {volume} {09}},\
  \bibinfo {pages} {030}},\ \Eprint {https://arxiv.org/abs/hep-ph/0006251}
  {arXiv:hep-ph/0006251 [hep-ph]} \BibitemShut {NoStop}%
\bibitem [{\citenamefont {Nelson}\ and\ \citenamefont
  {Strassler}(2002)}]{Nelson:2001mq}%
  \BibitemOpen
  \bibfield  {author} {\bibinfo {author} {\bibfnamefont {A.~E.}\ \bibnamefont
  {Nelson}}\ and\ \bibinfo {author} {\bibfnamefont {M.~J.}\ \bibnamefont
  {Strassler}},\ }\bibfield  {title} {\bibinfo {title} {{Exact results for
  supersymmetric renormalization and the supersymmetric flavor problem}},\
  }\href {https://doi.org/10.1088/1126-6708/2002/07/021} {\bibfield  {journal}
  {\bibinfo  {journal} {JHEP}\ }\textbf {\bibinfo {volume} {07}},\ \bibinfo
  {pages} {021}},\ \Eprint {https://arxiv.org/abs/hep-ph/0104051}
  {arXiv:hep-ph/0104051 [hep-ph]} \BibitemShut {NoStop}%
\bibitem [{\citenamefont {Schmaltz}\ and\ \citenamefont
  {Sundrum}(2006)}]{Schmaltz:2006qs}%
  \BibitemOpen
  \bibfield  {author} {\bibinfo {author} {\bibfnamefont {M.}~\bibnamefont
  {Schmaltz}}\ and\ \bibinfo {author} {\bibfnamefont {R.}~\bibnamefont
  {Sundrum}},\ }\bibfield  {title} {\bibinfo {title} {{Conformal Sequestering
  Simplified}},\ }\href {https://doi.org/10.1088/1126-6708/2006/11/011}
  {\bibfield  {journal} {\bibinfo  {journal} {JHEP}\ }\textbf {\bibinfo
  {volume} {11}},\ \bibinfo {pages} {011}},\ \Eprint
  {https://arxiv.org/abs/hep-th/0608051} {arXiv:hep-th/0608051 [hep-th]}
  \BibitemShut {NoStop}%
\bibitem [{\citenamefont {Choi}\ and\ \citenamefont
  {Jeong}(2007)}]{Choi:2006za}%
  \BibitemOpen
  \bibfield  {author} {\bibinfo {author} {\bibfnamefont {K.}~\bibnamefont
  {Choi}}\ and\ \bibinfo {author} {\bibfnamefont {K.~S.}\ \bibnamefont
  {Jeong}},\ }\bibfield  {title} {\bibinfo {title} {{String theoretic QCD axion
  with stabilized saxion and the pattern of supersymmetry breaking}},\ }\href
  {https://doi.org/10.1088/1126-6708/2007/01/103} {\bibfield  {journal}
  {\bibinfo  {journal} {JHEP}\ }\textbf {\bibinfo {volume} {01}},\ \bibinfo
  {pages} {103}},\ \Eprint {https://arxiv.org/abs/hep-th/0611279}
  {arXiv:hep-th/0611279 [hep-th]} \BibitemShut {NoStop}%
\bibitem [{\citenamefont {Kachru}\ \emph {et~al.}(2007)\citenamefont {Kachru},
  \citenamefont {McAllister},\ and\ \citenamefont {Sundrum}}]{Kachru:2007xp}%
  \BibitemOpen
  \bibfield  {author} {\bibinfo {author} {\bibfnamefont {S.}~\bibnamefont
  {Kachru}}, \bibinfo {author} {\bibfnamefont {L.}~\bibnamefont {McAllister}},\
  and\ \bibinfo {author} {\bibfnamefont {R.}~\bibnamefont {Sundrum}},\
  }\bibfield  {title} {\bibinfo {title} {{Sequestering in String Theory}},\
  }\href {https://doi.org/10.1088/1126-6708/2007/10/013} {\bibfield  {journal}
  {\bibinfo  {journal} {JHEP}\ }\textbf {\bibinfo {volume} {10}},\ \bibinfo
  {pages} {013}},\ \Eprint {https://arxiv.org/abs/hep-th/0703105}
  {arXiv:hep-th/0703105 [HEP-TH]} \BibitemShut {NoStop}%
\bibitem [{\citenamefont {Conlon}(2008)}]{Conlon:2007dw}%
  \BibitemOpen
  \bibfield  {author} {\bibinfo {author} {\bibfnamefont {J.~P.}\ \bibnamefont
  {Conlon}},\ }\bibfield  {title} {\bibinfo {title} {{Mirror Mediation}},\
  }\href {https://doi.org/10.1088/1126-6708/2008/03/025} {\bibfield  {journal}
  {\bibinfo  {journal} {JHEP}\ }\textbf {\bibinfo {volume} {03}},\ \bibinfo
  {pages} {025}},\ \Eprint {https://arxiv.org/abs/0710.0873} {arXiv:0710.0873
  [hep-th]} \BibitemShut {NoStop}%
\bibitem [{\citenamefont {Choi}\ \emph {et~al.}(2008)\citenamefont {Choi},
  \citenamefont {Jeong},\ and\ \citenamefont {Okumura}}]{Choi:2008hn}%
  \BibitemOpen
  \bibfield  {author} {\bibinfo {author} {\bibfnamefont {K.}~\bibnamefont
  {Choi}}, \bibinfo {author} {\bibfnamefont {K.~S.}\ \bibnamefont {Jeong}},\
  and\ \bibinfo {author} {\bibfnamefont {K.-i.}\ \bibnamefont {Okumura}},\
  }\bibfield  {title} {\bibinfo {title} {{Flavor and CP conserving moduli
  mediated SUSY breaking in flux compactification}},\ }\href
  {https://doi.org/10.1088/1126-6708/2008/07/047} {\bibfield  {journal}
  {\bibinfo  {journal} {JHEP}\ }\textbf {\bibinfo {volume} {07}},\ \bibinfo
  {pages} {047}},\ \Eprint {https://arxiv.org/abs/0804.4283} {arXiv:0804.4283
  [hep-ph]} \BibitemShut {NoStop}%
\bibitem [{\citenamefont {Inoue}\ \emph {et~al.}(1982)\citenamefont {Inoue},
  \citenamefont {Kakuto}, \citenamefont {Komatsu},\ and\ \citenamefont
  {Takeshita}}]{Inoue:1982pi}%
  \BibitemOpen
  \bibfield  {author} {\bibinfo {author} {\bibfnamefont {K.}~\bibnamefont
  {Inoue}}, \bibinfo {author} {\bibfnamefont {A.}~\bibnamefont {Kakuto}},
  \bibinfo {author} {\bibfnamefont {H.}~\bibnamefont {Komatsu}},\ and\ \bibinfo
  {author} {\bibfnamefont {S.}~\bibnamefont {Takeshita}},\ }\bibfield  {title}
  {\bibinfo {title} {{Aspects of Grand Unified Models with Softly Broken
  Supersymmetry}},\ }\href {https://doi.org/10.1143/PTP.68.927} {\bibfield
  {journal} {\bibinfo  {journal} {Prog. Theor. Phys.}\ }\textbf {\bibinfo
  {volume} {68}},\ \bibinfo {pages} {927} (\bibinfo {year} {1982})},\ \bibinfo
  {note} {[Erratum: Prog. Theor. Phys.70,330(1983)]}\BibitemShut {NoStop}%
\bibitem [{\citenamefont {Inoue}\ \emph {et~al.}(1984)\citenamefont {Inoue},
  \citenamefont {Kakuto}, \citenamefont {Komatsu},\ and\ \citenamefont
  {Takeshita}}]{Inoue:1983pp}%
  \BibitemOpen
  \bibfield  {author} {\bibinfo {author} {\bibfnamefont {K.}~\bibnamefont
  {Inoue}}, \bibinfo {author} {\bibfnamefont {A.}~\bibnamefont {Kakuto}},
  \bibinfo {author} {\bibfnamefont {H.}~\bibnamefont {Komatsu}},\ and\ \bibinfo
  {author} {\bibfnamefont {S.}~\bibnamefont {Takeshita}},\ }\bibfield  {title}
  {\bibinfo {title} {{Renormalization of Supersymmetry Breaking Parameters
  Revisited}},\ }\href {https://doi.org/10.1143/PTP.71.413} {\bibfield
  {journal} {\bibinfo  {journal} {Prog. Theor. Phys.}\ }\textbf {\bibinfo
  {volume} {71}},\ \bibinfo {pages} {413} (\bibinfo {year} {1984})}\BibitemShut
  {NoStop}%
\bibitem [{\citenamefont {Ibanez}\ and\ \citenamefont
  {Ross}(1982)}]{Ibanez:1982fr}%
  \BibitemOpen
  \bibfield  {author} {\bibinfo {author} {\bibfnamefont {L.~E.}\ \bibnamefont
  {Ibanez}}\ and\ \bibinfo {author} {\bibfnamefont {G.~G.}\ \bibnamefont
  {Ross}},\ }\bibfield  {title} {\bibinfo {title} {{SU(2)-L x U(1) Symmetry
  Breaking as a Radiative Effect of Supersymmetry Breaking in Guts}},\ }\href
  {https://doi.org/10.1016/0370-2693(82)91239-4} {\bibfield  {journal}
  {\bibinfo  {journal} {Phys. Lett.}\ }\textbf {\bibinfo {volume} {110B}},\
  \bibinfo {pages} {215} (\bibinfo {year} {1982})}\BibitemShut {NoStop}%
\bibitem [{\citenamefont {Ellis}\ \emph {et~al.}(1983)\citenamefont {Ellis},
  \citenamefont {Nanopoulos},\ and\ \citenamefont {Tamvakis}}]{Ellis:1982wr}%
  \BibitemOpen
  \bibfield  {author} {\bibinfo {author} {\bibfnamefont {J.~R.}\ \bibnamefont
  {Ellis}}, \bibinfo {author} {\bibfnamefont {D.~V.}\ \bibnamefont
  {Nanopoulos}},\ and\ \bibinfo {author} {\bibfnamefont {K.}~\bibnamefont
  {Tamvakis}},\ }\bibfield  {title} {\bibinfo {title} {{Grand Unification in
  Simple Supergravity}},\ }\href {https://doi.org/10.1016/0370-2693(83)90900-0}
  {\bibfield  {journal} {\bibinfo  {journal} {Phys. Lett.}\ }\textbf {\bibinfo
  {volume} {121B}},\ \bibinfo {pages} {123} (\bibinfo {year}
  {1983})}\BibitemShut {NoStop}%
\bibitem [{\citenamefont {Alvarez-Gaume}\ \emph {et~al.}(1983)\citenamefont
  {Alvarez-Gaume}, \citenamefont {Polchinski},\ and\ \citenamefont
  {Wise}}]{AlvarezGaume:1983gj}%
  \BibitemOpen
  \bibfield  {author} {\bibinfo {author} {\bibfnamefont {L.}~\bibnamefont
  {Alvarez-Gaume}}, \bibinfo {author} {\bibfnamefont {J.}~\bibnamefont
  {Polchinski}},\ and\ \bibinfo {author} {\bibfnamefont {M.~B.}\ \bibnamefont
  {Wise}},\ }\bibfield  {title} {\bibinfo {title} {{Minimal Low-Energy
  Supergravity}},\ }\href {https://doi.org/10.1016/0550-3213(83)90591-6}
  {\bibfield  {journal} {\bibinfo  {journal} {Nucl. Phys.}\ }\textbf {\bibinfo
  {volume} {B221}},\ \bibinfo {pages} {495} (\bibinfo {year}
  {1983})}\BibitemShut {NoStop}%
\bibitem [{\citenamefont {Barbieri}\ and\ \citenamefont
  {Hall}(1994)}]{Barbieri:1994pv}%
  \BibitemOpen
  \bibfield  {author} {\bibinfo {author} {\bibfnamefont {R.}~\bibnamefont
  {Barbieri}}\ and\ \bibinfo {author} {\bibfnamefont {L.~J.}\ \bibnamefont
  {Hall}},\ }\bibfield  {title} {\bibinfo {title} {{Signals for supersymmetric
  unification}},\ }\href {https://doi.org/10.1016/0370-2693(94)91368-4}
  {\bibfield  {journal} {\bibinfo  {journal} {Phys. Lett.}\ }\textbf {\bibinfo
  {volume} {B338}},\ \bibinfo {pages} {212} (\bibinfo {year} {1994})},\ \Eprint
  {https://arxiv.org/abs/hep-ph/9408406} {arXiv:hep-ph/9408406 [hep-ph]}
  \BibitemShut {NoStop}%
\bibitem [{\citenamefont {Barbieri}\ \emph {et~al.}(1995)\citenamefont
  {Barbieri}, \citenamefont {Hall},\ and\ \citenamefont
  {Strumia}}]{Barbieri:1995tw}%
  \BibitemOpen
  \bibfield  {author} {\bibinfo {author} {\bibfnamefont {R.}~\bibnamefont
  {Barbieri}}, \bibinfo {author} {\bibfnamefont {L.~J.}\ \bibnamefont {Hall}},\
  and\ \bibinfo {author} {\bibfnamefont {A.}~\bibnamefont {Strumia}},\
  }\bibfield  {title} {\bibinfo {title} {{Violations of lepton flavor and CP in
  supersymmetric unified theories}},\ }\href
  {https://doi.org/10.1016/0550-3213(95)00208-A} {\bibfield  {journal}
  {\bibinfo  {journal} {Nucl. Phys.}\ }\textbf {\bibinfo {volume} {B445}},\
  \bibinfo {pages} {219} (\bibinfo {year} {1995})},\ \Eprint
  {https://arxiv.org/abs/hep-ph/9501334} {arXiv:hep-ph/9501334 [hep-ph]}
  \BibitemShut {NoStop}%
\bibitem [{\citenamefont {Borzumati}\ and\ \citenamefont
  {Masiero}(1986)}]{Borzumati:1986qx}%
  \BibitemOpen
  \bibfield  {author} {\bibinfo {author} {\bibfnamefont {F.}~\bibnamefont
  {Borzumati}}\ and\ \bibinfo {author} {\bibfnamefont {A.}~\bibnamefont
  {Masiero}},\ }\bibfield  {title} {\bibinfo {title} {{Large Muon and electron
  Number Violations in Supergravity Theories}},\ }\href
  {https://doi.org/10.1103/PhysRevLett.57.961} {\bibfield  {journal} {\bibinfo
  {journal} {Phys. Rev. Lett.}\ }\textbf {\bibinfo {volume} {57}},\ \bibinfo
  {pages} {961} (\bibinfo {year} {1986})}\BibitemShut {NoStop}%
\bibitem [{\citenamefont {Hisano}\ \emph {et~al.}(1995)\citenamefont {Hisano},
  \citenamefont {Moroi}, \citenamefont {Tobe}, \citenamefont {Yamaguchi},\ and\
  \citenamefont {Yanagida}}]{Hisano:1995nq}%
  \BibitemOpen
  \bibfield  {author} {\bibinfo {author} {\bibfnamefont {J.}~\bibnamefont
  {Hisano}}, \bibinfo {author} {\bibfnamefont {T.}~\bibnamefont {Moroi}},
  \bibinfo {author} {\bibfnamefont {K.}~\bibnamefont {Tobe}}, \bibinfo {author}
  {\bibfnamefont {M.}~\bibnamefont {Yamaguchi}},\ and\ \bibinfo {author}
  {\bibfnamefont {T.}~\bibnamefont {Yanagida}},\ }\bibfield  {title} {\bibinfo
  {title} {{Lepton flavor violation in the supersymmetric standard model with
  seesaw induced neutrino masses}},\ }\href
  {https://doi.org/10.1016/0370-2693(95)00954-J} {\bibfield  {journal}
  {\bibinfo  {journal} {Phys. Lett.}\ }\textbf {\bibinfo {volume} {B357}},\
  \bibinfo {pages} {579} (\bibinfo {year} {1995})},\ \Eprint
  {https://arxiv.org/abs/hep-ph/9501407} {arXiv:hep-ph/9501407 [hep-ph]}
  \BibitemShut {NoStop}%
\bibitem [{\citenamefont {Hisano}\ \emph {et~al.}(1996)\citenamefont {Hisano},
  \citenamefont {Moroi}, \citenamefont {Tobe},\ and\ \citenamefont
  {Yamaguchi}}]{Hisano:1995cp}%
  \BibitemOpen
  \bibfield  {author} {\bibinfo {author} {\bibfnamefont {J.}~\bibnamefont
  {Hisano}}, \bibinfo {author} {\bibfnamefont {T.}~\bibnamefont {Moroi}},
  \bibinfo {author} {\bibfnamefont {K.}~\bibnamefont {Tobe}},\ and\ \bibinfo
  {author} {\bibfnamefont {M.}~\bibnamefont {Yamaguchi}},\ }\bibfield  {title}
  {\bibinfo {title} {{Lepton flavor violation via right-handed neutrino Yukawa
  couplings in supersymmetric standard model}},\ }\href
  {https://doi.org/10.1103/PhysRevD.53.2442} {\bibfield  {journal} {\bibinfo
  {journal} {Phys. Rev.}\ }\textbf {\bibinfo {volume} {D53}},\ \bibinfo {pages}
  {2442} (\bibinfo {year} {1996})},\ \Eprint
  {https://arxiv.org/abs/hep-ph/9510309} {arXiv:hep-ph/9510309 [hep-ph]}
  \BibitemShut {NoStop}%
\bibitem [{\citenamefont {Hisano}\ \emph {et~al.}(1998)\citenamefont {Hisano},
  \citenamefont {Nomura},\ and\ \citenamefont {Yanagida}}]{Hisano:1997tc}%
  \BibitemOpen
  \bibfield  {author} {\bibinfo {author} {\bibfnamefont {J.}~\bibnamefont
  {Hisano}}, \bibinfo {author} {\bibfnamefont {D.}~\bibnamefont {Nomura}},\
  and\ \bibinfo {author} {\bibfnamefont {T.}~\bibnamefont {Yanagida}},\
  }\bibfield  {title} {\bibinfo {title} {{Atmospheric neutrino oscillation and
  large lepton flavor violation in the SUSY SU(5) GUT}},\ }\href
  {https://doi.org/10.1016/S0370-2693(98)00929-0} {\bibfield  {journal}
  {\bibinfo  {journal} {Phys. Lett.}\ }\textbf {\bibinfo {volume} {B437}},\
  \bibinfo {pages} {351} (\bibinfo {year} {1998})},\ \Eprint
  {https://arxiv.org/abs/hep-ph/9711348} {arXiv:hep-ph/9711348 [hep-ph]}
  \BibitemShut {NoStop}%
\bibitem [{\citenamefont {Hisano}\ and\ \citenamefont
  {Nomura}(1999)}]{Hisano:1998fj}%
  \BibitemOpen
  \bibfield  {author} {\bibinfo {author} {\bibfnamefont {J.}~\bibnamefont
  {Hisano}}\ and\ \bibinfo {author} {\bibfnamefont {D.}~\bibnamefont
  {Nomura}},\ }\bibfield  {title} {\bibinfo {title} {{Solar and atmospheric
  neutrino oscillations and lepton flavor violation in supersymmetric models
  with the right-handed neutrinos}},\ }\href
  {https://doi.org/10.1103/PhysRevD.59.116005} {\bibfield  {journal} {\bibinfo
  {journal} {Phys. Rev.}\ }\textbf {\bibinfo {volume} {D59}},\ \bibinfo {pages}
  {116005} (\bibinfo {year} {1999})},\ \Eprint
  {https://arxiv.org/abs/hep-ph/9810479} {arXiv:hep-ph/9810479 [hep-ph]}
  \BibitemShut {NoStop}%
\bibitem [{\citenamefont {Baek}\ \emph
  {et~al.}(2001{\natexlab{a}})\citenamefont {Baek}, \citenamefont {Goto},
  \citenamefont {Okada},\ and\ \citenamefont {Okumura}}]{Baek:2000sj}%
  \BibitemOpen
  \bibfield  {author} {\bibinfo {author} {\bibfnamefont {S.}~\bibnamefont
  {Baek}}, \bibinfo {author} {\bibfnamefont {T.}~\bibnamefont {Goto}}, \bibinfo
  {author} {\bibfnamefont {Y.}~\bibnamefont {Okada}},\ and\ \bibinfo {author}
  {\bibfnamefont {K.-i.}\ \bibnamefont {Okumura}},\ }\bibfield  {title}
  {\bibinfo {title} {{Neutrino oscillation, SUSY GUT and B decay}},\ }\href
  {https://doi.org/10.1103/PhysRevD.63.051701} {\bibfield  {journal} {\bibinfo
  {journal} {Phys. Rev.}\ }\textbf {\bibinfo {volume} {D63}},\ \bibinfo {pages}
  {051701} (\bibinfo {year} {2001}{\natexlab{a}})},\ \Eprint
  {https://arxiv.org/abs/hep-ph/0002141} {arXiv:hep-ph/0002141 [hep-ph]}
  \BibitemShut {NoStop}%
\bibitem [{\citenamefont {Moroi}(2000)}]{Moroi:2000mr}%
  \BibitemOpen
  \bibfield  {author} {\bibinfo {author} {\bibfnamefont {T.}~\bibnamefont
  {Moroi}},\ }\bibfield  {title} {\bibinfo {title} {{Effects of the
  right-handed neutrinos on Delta S = 2 and Delta B = 2 processes in
  supersymmetric SU(5) model}},\ }\href
  {https://doi.org/10.1088/1126-6708/2000/03/019} {\bibfield  {journal}
  {\bibinfo  {journal} {JHEP}\ }\textbf {\bibinfo {volume} {03}},\ \bibinfo
  {pages} {019}},\ \Eprint {https://arxiv.org/abs/hep-ph/0002208}
  {arXiv:hep-ph/0002208 [hep-ph]} \BibitemShut {NoStop}%
\bibitem [{\citenamefont {Baek}\ \emph
  {et~al.}(2001{\natexlab{b}})\citenamefont {Baek}, \citenamefont {Goto},
  \citenamefont {Okada},\ and\ \citenamefont {Okumura}}]{Baek:2001kh}%
  \BibitemOpen
  \bibfield  {author} {\bibinfo {author} {\bibfnamefont {S.}~\bibnamefont
  {Baek}}, \bibinfo {author} {\bibfnamefont {T.}~\bibnamefont {Goto}}, \bibinfo
  {author} {\bibfnamefont {Y.}~\bibnamefont {Okada}},\ and\ \bibinfo {author}
  {\bibfnamefont {K.-i.}\ \bibnamefont {Okumura}},\ }\bibfield  {title}
  {\bibinfo {title} {{Muon anomalous magnetic moment, lepton flavor violation,
  and flavor changing neutral current processes in SUSY GUT with right-handed
  neutrino}},\ }\href {https://doi.org/10.1103/PhysRevD.64.095001} {\bibfield
  {journal} {\bibinfo  {journal} {Phys. Rev.}\ }\textbf {\bibinfo {volume}
  {D64}},\ \bibinfo {pages} {095001} (\bibinfo {year} {2001}{\natexlab{b}})},\
  \Eprint {https://arxiv.org/abs/hep-ph/0104146} {arXiv:hep-ph/0104146
  [hep-ph]} \BibitemShut {NoStop}%
\bibitem [{Note1()}]{Note1}%
  \BibitemOpen
  \bibinfo {note} {We follow the discussion in \cite {Choi:2005ge,
  Choi:2005uz}.}\BibitemShut {Stop}%
\bibitem [{\citenamefont {Wess}\ and\ \citenamefont {Bagger}(1992)}]{Wess92}%
  \BibitemOpen
  \bibfield  {author} {\bibinfo {author} {\bibfnamefont {J.}~\bibnamefont
  {Wess}}\ and\ \bibinfo {author} {\bibfnamefont {J.}~\bibnamefont {Bagger}},\
  }\href@noop {} {\emph {\bibinfo {title} {Supersymmetry and Supergravity}}}\
  (\bibinfo  {publisher} {Princeton University Press},\ \bibinfo {year}
  {1992})\BibitemShut {NoStop}%
\bibitem [{\citenamefont {Kachru}\ \emph {et~al.}(2003)\citenamefont {Kachru},
  \citenamefont {Kallosh}, \citenamefont {Linde},\ and\ \citenamefont
  {Trivedi}}]{Kachru:2003aw}%
  \BibitemOpen
  \bibfield  {author} {\bibinfo {author} {\bibfnamefont {S.}~\bibnamefont
  {Kachru}}, \bibinfo {author} {\bibfnamefont {R.}~\bibnamefont {Kallosh}},
  \bibinfo {author} {\bibfnamefont {A.~D.}\ \bibnamefont {Linde}},\ and\
  \bibinfo {author} {\bibfnamefont {S.~P.}\ \bibnamefont {Trivedi}},\
  }\bibfield  {title} {\bibinfo {title} {{De Sitter vacua in string theory}},\
  }\href {https://doi.org/10.1103/PhysRevD.68.046005} {\bibfield  {journal}
  {\bibinfo  {journal} {Phys. Rev.}\ }\textbf {\bibinfo {volume} {D68}},\
  \bibinfo {pages} {046005} (\bibinfo {year} {2003})},\ \Eprint
  {https://arxiv.org/abs/hep-th/0301240} {arXiv:hep-th/0301240 [hep-th]}
  \BibitemShut {NoStop}%
\bibitem [{\citenamefont {Choi}\ \emph
  {et~al.}(2005{\natexlab{a}})\citenamefont {Choi}, \citenamefont {Falkowski},
  \citenamefont {Nilles},\ and\ \citenamefont {Olechowski}}]{Choi:2005ge}%
  \BibitemOpen
  \bibfield  {author} {\bibinfo {author} {\bibfnamefont {K.}~\bibnamefont
  {Choi}}, \bibinfo {author} {\bibfnamefont {A.}~\bibnamefont {Falkowski}},
  \bibinfo {author} {\bibfnamefont {H.~P.}\ \bibnamefont {Nilles}},\ and\
  \bibinfo {author} {\bibfnamefont {M.}~\bibnamefont {Olechowski}},\ }\bibfield
   {title} {\bibinfo {title} {{Soft supersymmetry breaking in KKLT flux
  compactification}},\ }\href {https://doi.org/10.1016/j.nuclphysb.2005.04.032}
  {\bibfield  {journal} {\bibinfo  {journal} {Nucl. Phys.}\ }\textbf {\bibinfo
  {volume} {B718}},\ \bibinfo {pages} {113} (\bibinfo {year}
  {2005}{\natexlab{a}})},\ \Eprint {https://arxiv.org/abs/hep-th/0503216}
  {arXiv:hep-th/0503216 [hep-th]} \BibitemShut {NoStop}%
\bibitem [{\citenamefont {Choi}\ \emph
  {et~al.}(2005{\natexlab{b}})\citenamefont {Choi}, \citenamefont {Jeong},\
  and\ \citenamefont {Okumura}}]{Choi:2005uz}%
  \BibitemOpen
  \bibfield  {author} {\bibinfo {author} {\bibfnamefont {K.}~\bibnamefont
  {Choi}}, \bibinfo {author} {\bibfnamefont {K.~S.}\ \bibnamefont {Jeong}},\
  and\ \bibinfo {author} {\bibfnamefont {K.-i.}\ \bibnamefont {Okumura}},\
  }\bibfield  {title} {\bibinfo {title} {{Phenomenology of mixed
  modulus-anomaly mediation in fluxed string compactifications and brane
  models}},\ }\href {https://doi.org/10.1088/1126-6708/2005/09/039} {\bibfield
  {journal} {\bibinfo  {journal} {JHEP}\ }\textbf {\bibinfo {volume} {09}},\
  \bibinfo {pages} {039}},\ \Eprint {https://arxiv.org/abs/hep-ph/0504037}
  {arXiv:hep-ph/0504037 [hep-ph]} \BibitemShut {NoStop}%
\bibitem [{\citenamefont {Choi}\ \emph {et~al.}(2004)\citenamefont {Choi},
  \citenamefont {Falkowski}, \citenamefont {Nilles}, \citenamefont
  {Olechowski},\ and\ \citenamefont {Pokorski}}]{Choi:2004sx}%
  \BibitemOpen
  \bibfield  {author} {\bibinfo {author} {\bibfnamefont {K.}~\bibnamefont
  {Choi}}, \bibinfo {author} {\bibfnamefont {A.}~\bibnamefont {Falkowski}},
  \bibinfo {author} {\bibfnamefont {H.~P.}\ \bibnamefont {Nilles}}, \bibinfo
  {author} {\bibfnamefont {M.}~\bibnamefont {Olechowski}},\ and\ \bibinfo
  {author} {\bibfnamefont {S.}~\bibnamefont {Pokorski}},\ }\bibfield  {title}
  {\bibinfo {title} {{Stability of flux compactifications and the pattern of
  supersymmetry breaking}},\ }\href
  {https://doi.org/10.1088/1126-6708/2004/11/076} {\bibfield  {journal}
  {\bibinfo  {journal} {JHEP}\ }\textbf {\bibinfo {volume} {11}},\ \bibinfo
  {pages} {076}},\ \Eprint {https://arxiv.org/abs/hep-th/0411066}
  {arXiv:hep-th/0411066 [hep-th]} \BibitemShut {NoStop}%
\bibitem [{\citenamefont {Choi}\ \emph {et~al.}(2006)\citenamefont {Choi},
  \citenamefont {Jeong}, \citenamefont {Kobayashi},\ and\ \citenamefont
  {Okumura}}]{Choi:2005hd}%
  \BibitemOpen
  \bibfield  {author} {\bibinfo {author} {\bibfnamefont {K.}~\bibnamefont
  {Choi}}, \bibinfo {author} {\bibfnamefont {K.~S.}\ \bibnamefont {Jeong}},
  \bibinfo {author} {\bibfnamefont {T.}~\bibnamefont {Kobayashi}},\ and\
  \bibinfo {author} {\bibfnamefont {K.-i.}\ \bibnamefont {Okumura}},\
  }\bibfield  {title} {\bibinfo {title} {{Little SUSY hierarchy in mixed
  modulus-anomaly mediation}},\ }\href
  {https://doi.org/10.1016/j.physletb.2005.11.078} {\bibfield  {journal}
  {\bibinfo  {journal} {Phys. Lett.}\ }\textbf {\bibinfo {volume} {B633}},\
  \bibinfo {pages} {355} (\bibinfo {year} {2006})},\ \Eprint
  {https://arxiv.org/abs/hep-ph/0508029} {arXiv:hep-ph/0508029 [hep-ph]}
  \BibitemShut {NoStop}%
\bibitem [{\citenamefont {Endo}\ \emph {et~al.}(2005)\citenamefont {Endo},
  \citenamefont {Yamaguchi},\ and\ \citenamefont {Yoshioka}}]{Endo:2005uy}%
  \BibitemOpen
  \bibfield  {author} {\bibinfo {author} {\bibfnamefont {M.}~\bibnamefont
  {Endo}}, \bibinfo {author} {\bibfnamefont {M.}~\bibnamefont {Yamaguchi}},\
  and\ \bibinfo {author} {\bibfnamefont {K.}~\bibnamefont {Yoshioka}},\
  }\bibfield  {title} {\bibinfo {title} {{A Bottom-up approach to moduli
  dynamics in heavy gravitino scenario: Superpotential, soft terms and
  sparticle mass spectrum}},\ }\href
  {https://doi.org/10.1103/PhysRevD.72.015004} {\bibfield  {journal} {\bibinfo
  {journal} {Phys. Rev.}\ }\textbf {\bibinfo {volume} {D72}},\ \bibinfo {pages}
  {015004} (\bibinfo {year} {2005})},\ \Eprint
  {https://arxiv.org/abs/hep-ph/0504036} {arXiv:hep-ph/0504036 [hep-ph]}
  \BibitemShut {NoStop}%
\bibitem [{\citenamefont {Arnold}(1978)}]{Arnold78}%
  \BibitemOpen
  \bibfield  {author} {\bibinfo {author} {\bibfnamefont {V.~I.}\ \bibnamefont
  {Arnold}},\ }\href@noop {} {\emph {\bibinfo {title} {Mathematical Methods of
  Classical Mechanics}}}\ (\bibinfo  {publisher} {Springer},\ \bibinfo {year}
  {1978})\BibitemShut {NoStop}%
\bibitem [{\citenamefont {Landau}\ and\ \citenamefont
  {Lifshitz}(1976)}]{Landau76}%
  \BibitemOpen
  \bibfield  {author} {\bibinfo {author} {\bibfnamefont {L.~D.}\ \bibnamefont
  {Landau}}\ and\ \bibinfo {author} {\bibfnamefont {E.~M.}\ \bibnamefont
  {Lifshitz}},\ }\href@noop {} {\emph {\bibinfo {title} {Mechanics}}}\
  (\bibinfo  {publisher} {Elsevier},\ \bibinfo {year} {1976})\BibitemShut
  {NoStop}%
\bibitem [{\citenamefont {Kitano}\ and\ \citenamefont
  {Nomura}(2006)}]{Kitano:2006gv}%
  \BibitemOpen
  \bibfield  {author} {\bibinfo {author} {\bibfnamefont {R.}~\bibnamefont
  {Kitano}}\ and\ \bibinfo {author} {\bibfnamefont {Y.}~\bibnamefont
  {Nomura}},\ }\bibfield  {title} {\bibinfo {title} {{Supersymmetry,
  naturalness, and signatures at the LHC}},\ }\href
  {https://doi.org/10.1103/PhysRevD.73.095004} {\bibfield  {journal} {\bibinfo
  {journal} {Phys. Rev.}\ }\textbf {\bibinfo {volume} {D73}},\ \bibinfo {pages}
  {095004} (\bibinfo {year} {2006})},\ \Eprint
  {https://arxiv.org/abs/hep-ph/0602096} {arXiv:hep-ph/0602096 [hep-ph]}
  \BibitemShut {NoStop}%
\bibitem [{\citenamefont {Burgess}\ \emph {et~al.}(1986)\citenamefont
  {Burgess}, \citenamefont {Font},\ and\ \citenamefont
  {Quevedo}}]{Burgess:1985zz}%
  \BibitemOpen
  \bibfield  {author} {\bibinfo {author} {\bibfnamefont {C.~P.}\ \bibnamefont
  {Burgess}}, \bibinfo {author} {\bibfnamefont {A.}~\bibnamefont {Font}},\ and\
  \bibinfo {author} {\bibfnamefont {F.}~\bibnamefont {Quevedo}},\ }\bibfield
  {title} {\bibinfo {title} {{Low-Energy Effective Action for the
  Superstring}},\ }\href {https://doi.org/10.1016/0550-3213(86)90239-7}
  {\bibfield  {journal} {\bibinfo  {journal} {Nucl. Phys.}\ }\textbf {\bibinfo
  {volume} {B272}},\ \bibinfo {pages} {661} (\bibinfo {year}
  {1986})}\BibitemShut {NoStop}%
\bibitem [{\citenamefont {Nilles}(1986)}]{Nilles:1986cy}%
  \BibitemOpen
  \bibfield  {author} {\bibinfo {author} {\bibfnamefont {H.~P.}\ \bibnamefont
  {Nilles}},\ }\bibfield  {title} {\bibinfo {title} {{The Role of Classical
  Symmetries in the Low-energy Limit of Superstring Theories}},\ }\href
  {https://doi.org/10.1016/0370-2693(86)90302-3} {\bibfield  {journal}
  {\bibinfo  {journal} {Phys. Lett.}\ }\textbf {\bibinfo {volume} {B180}},\
  \bibinfo {pages} {240} (\bibinfo {year} {1986})}\BibitemShut {NoStop}%
\bibitem [{\citenamefont {Choi}(1988)}]{Choi:1987is}%
  \BibitemOpen
  \bibfield  {author} {\bibinfo {author} {\bibfnamefont {K.}~\bibnamefont
  {Choi}},\ }\bibfield  {title} {\bibinfo {title} {{Supersymmetry Breaking for
  the Observable Sector in Superstring Models}},\ }\href
  {https://doi.org/10.1007/BF01550997} {\bibfield  {journal} {\bibinfo
  {journal} {Z. Phys.}\ }\textbf {\bibinfo {volume} {C39}},\ \bibinfo {pages}
  {219} (\bibinfo {year} {1988})}\BibitemShut {NoStop}%
\bibitem [{\citenamefont {Conlon}\ \emph {et~al.}(2007)\citenamefont {Conlon},
  \citenamefont {Cremades},\ and\ \citenamefont {Quevedo}}]{Conlon:2006tj}%
  \BibitemOpen
  \bibfield  {author} {\bibinfo {author} {\bibfnamefont {J.~P.}\ \bibnamefont
  {Conlon}}, \bibinfo {author} {\bibfnamefont {D.}~\bibnamefont {Cremades}},\
  and\ \bibinfo {author} {\bibfnamefont {F.}~\bibnamefont {Quevedo}},\
  }\bibfield  {title} {\bibinfo {title} {{Kahler potentials of chiral matter
  fields for Calabi-Yau string compactifications}},\ }\href
  {https://doi.org/10.1088/1126-6708/2007/01/022} {\bibfield  {journal}
  {\bibinfo  {journal} {JHEP}\ }\textbf {\bibinfo {volume} {01}},\ \bibinfo
  {pages} {022}},\ \Eprint {https://arxiv.org/abs/hep-th/0609180}
  {arXiv:hep-th/0609180 [hep-th]} \BibitemShut {NoStop}%
\bibitem [{Note2()}]{Note2}%
  \BibitemOpen
  \bibinfo {note} {Precisely, the threshold is given by $X/\protect \sqrt
  {Z_{\Psi ^1}\protect \,Z_{\Psi ^2}}$. The difference in the estimate of the
  SUSY breaking appears to be the next-to-leading logarithm order since picking
  up a $F$ component in $Z_{\Psi ^\alpha }$ eliminates one
  logarithm.}\BibitemShut {Stop}%
\bibitem [{\citenamefont {Kim}\ and\ \citenamefont
  {Nilles}(1991)}]{Kim:1991mv}%
  \BibitemOpen
  \bibfield  {author} {\bibinfo {author} {\bibfnamefont {J.~E.}\ \bibnamefont
  {Kim}}\ and\ \bibinfo {author} {\bibfnamefont {H.~P.}\ \bibnamefont
  {Nilles}},\ }\bibfield  {title} {\bibinfo {title} {{Gaugino condensation and
  the cosmological implications of the hidden sector}},\ }\href
  {https://doi.org/10.1016/0370-2693(91)91710-D} {\bibfield  {journal}
  {\bibinfo  {journal} {Phys. Lett.}\ }\textbf {\bibinfo {volume} {B263}},\
  \bibinfo {pages} {79} (\bibinfo {year} {1991})}\BibitemShut {NoStop}%
\bibitem [{\citenamefont {Chun}\ \emph {et~al.}(1992)\citenamefont {Chun},
  \citenamefont {Kim},\ and\ \citenamefont {Nilles}}]{Chun:1991xm}%
  \BibitemOpen
  \bibfield  {author} {\bibinfo {author} {\bibfnamefont {E.~J.}\ \bibnamefont
  {Chun}}, \bibinfo {author} {\bibfnamefont {J.~E.}\ \bibnamefont {Kim}},\ and\
  \bibinfo {author} {\bibfnamefont {H.~P.}\ \bibnamefont {Nilles}},\ }\bibfield
   {title} {\bibinfo {title} {{A Natural solution of the mu problem with a
  composite axion in the hidden sector}},\ }\href
  {https://doi.org/10.1016/0550-3213(92)90346-D} {\bibfield  {journal}
  {\bibinfo  {journal} {Nucl. Phys.}\ }\textbf {\bibinfo {volume} {B370}},\
  \bibinfo {pages} {105} (\bibinfo {year} {1992})}\BibitemShut {NoStop}%
\bibitem [{\citenamefont {Faraggi}\ and\ \citenamefont
  {Halyo}(1993)}]{Faraggi:1993zh}%
  \BibitemOpen
  \bibfield  {author} {\bibinfo {author} {\bibfnamefont {A.~E.}\ \bibnamefont
  {Faraggi}}\ and\ \bibinfo {author} {\bibfnamefont {E.}~\bibnamefont
  {Halyo}},\ }\bibfield  {title} {\bibinfo {title} {{Neutrino masses in
  superstring derived standard - like models}},\ }\href
  {https://doi.org/10.1016/0370-2693(93)90226-8} {\bibfield  {journal}
  {\bibinfo  {journal} {Phys. Lett.}\ }\textbf {\bibinfo {volume} {B307}},\
  \bibinfo {pages} {311} (\bibinfo {year} {1993})},\ \Eprint
  {https://arxiv.org/abs/hep-th/9303060} {arXiv:hep-th/9303060 [hep-th]}
  \BibitemShut {NoStop}%
\bibitem [{\citenamefont {Coriano}\ and\ \citenamefont
  {Faraggi}(2004)}]{Coriano:2003ui}%
  \BibitemOpen
  \bibfield  {author} {\bibinfo {author} {\bibfnamefont {C.}~\bibnamefont
  {Coriano}}\ and\ \bibinfo {author} {\bibfnamefont {A.~E.}\ \bibnamefont
  {Faraggi}},\ }\bibfield  {title} {\bibinfo {title} {{String inspired neutrino
  mass textures in light of KamLAND and WMAP}},\ }\href
  {https://doi.org/10.1016/j.physletb.2003.11.071} {\bibfield  {journal}
  {\bibinfo  {journal} {Phys. Lett.}\ }\textbf {\bibinfo {volume} {B581}},\
  \bibinfo {pages} {99} (\bibinfo {year} {2004})},\ \Eprint
  {https://arxiv.org/abs/hep-ph/0306186} {arXiv:hep-ph/0306186 [hep-ph]}
  \BibitemShut {NoStop}%
\bibitem [{\citenamefont {Blumenhagen}\ \emph {et~al.}(2007)\citenamefont
  {Blumenhagen}, \citenamefont {Cvetic},\ and\ \citenamefont
  {Weigand}}]{Blumenhagen:2006xt}%
  \BibitemOpen
  \bibfield  {author} {\bibinfo {author} {\bibfnamefont {R.}~\bibnamefont
  {Blumenhagen}}, \bibinfo {author} {\bibfnamefont {M.}~\bibnamefont
  {Cvetic}},\ and\ \bibinfo {author} {\bibfnamefont {T.}~\bibnamefont
  {Weigand}},\ }\bibfield  {title} {\bibinfo {title} {{Spacetime instanton
  corrections in 4D string vacua: The Seesaw mechanism for D-Brane models}},\
  }\href {https://doi.org/10.1016/j.nuclphysb.2007.02.016} {\bibfield
  {journal} {\bibinfo  {journal} {Nucl. Phys.}\ }\textbf {\bibinfo {volume}
  {B771}},\ \bibinfo {pages} {113} (\bibinfo {year} {2007})},\ \Eprint
  {https://arxiv.org/abs/hep-th/0609191} {arXiv:hep-th/0609191 [hep-th]}
  \BibitemShut {NoStop}%
\bibitem [{\citenamefont {Ibanez}\ \emph {et~al.}(2007)\citenamefont {Ibanez},
  \citenamefont {Schellekens},\ and\ \citenamefont {Uranga}}]{Ibanez:2007rs}%
  \BibitemOpen
  \bibfield  {author} {\bibinfo {author} {\bibfnamefont {L.~E.}\ \bibnamefont
  {Ibanez}}, \bibinfo {author} {\bibfnamefont {A.~N.}\ \bibnamefont
  {Schellekens}},\ and\ \bibinfo {author} {\bibfnamefont {A.~M.}\ \bibnamefont
  {Uranga}},\ }\bibfield  {title} {\bibinfo {title} {{Instanton Induced
  Neutrino Majorana Masses in CFT Orientifolds with MSSM-like spectra}},\
  }\href {https://doi.org/10.1088/1126-6708/2007/06/011} {\bibfield  {journal}
  {\bibinfo  {journal} {JHEP}\ }\textbf {\bibinfo {volume} {06}},\ \bibinfo
  {pages} {011}},\ \Eprint {https://arxiv.org/abs/0704.1079} {arXiv:0704.1079
  [hep-th]} \BibitemShut {NoStop}%
\bibitem [{\citenamefont {Florea}\ \emph {et~al.}(2007)\citenamefont {Florea},
  \citenamefont {Kachru}, \citenamefont {McGreevy},\ and\ \citenamefont
  {Saulina}}]{Florea:2006si}%
  \BibitemOpen
  \bibfield  {author} {\bibinfo {author} {\bibfnamefont {B.}~\bibnamefont
  {Florea}}, \bibinfo {author} {\bibfnamefont {S.}~\bibnamefont {Kachru}},
  \bibinfo {author} {\bibfnamefont {J.}~\bibnamefont {McGreevy}},\ and\
  \bibinfo {author} {\bibfnamefont {N.}~\bibnamefont {Saulina}},\ }\bibfield
  {title} {\bibinfo {title} {{Stringy Instantons and Quiver Gauge Theories}},\
  }\href {https://doi.org/10.1088/1126-6708/2007/05/024} {\bibfield  {journal}
  {\bibinfo  {journal} {JHEP}\ }\textbf {\bibinfo {volume} {05}},\ \bibinfo
  {pages} {024}},\ \Eprint {https://arxiv.org/abs/hep-th/0610003}
  {arXiv:hep-th/0610003 [hep-th]} \BibitemShut {NoStop}%
\bibitem [{\citenamefont {Langacker}(2012)}]{Langacker:2011bi}%
  \BibitemOpen
  \bibfield  {author} {\bibinfo {author} {\bibfnamefont {P.}~\bibnamefont
  {Langacker}},\ }\bibfield  {title} {\bibinfo {title} {{Neutrino Masses from
  the Top Down}},\ }\href {https://doi.org/10.1146/annurev-nucl-102711-094925}
  {\bibfield  {journal} {\bibinfo  {journal} {Ann. Rev. Nucl. Part. Sci.}\
  }\textbf {\bibinfo {volume} {62}},\ \bibinfo {pages} {215} (\bibinfo {year}
  {2012})},\ \Eprint {https://arxiv.org/abs/1112.5992} {arXiv:1112.5992
  [hep-ph]} \BibitemShut {NoStop}%
\bibitem [{Note3()}]{Note3}%
  \BibitemOpen
  \bibinfo {note} {See also \cite
  {Pomarol:1999ie,Rattazzi:1999qg,Okada:2002mv,Everett:2008qy,Everett:2008ey,Choi:2009jn,Altunkaynak:2010xe}
  for studies on the threshold corrections in the modulus and anomaly
  mediation.}\BibitemShut {Stop}%
\bibitem [{\citenamefont {Yanagida}(1979)}]{Yanagida:1979as}%
  \BibitemOpen
  \bibfield  {author} {\bibinfo {author} {\bibfnamefont {T.}~\bibnamefont
  {Yanagida}},\ }\bibfield  {title} {\bibinfo {title} {{Horizontal gauge
  symmetry and masses of neutrinos}},\ }\bibfield  {booktitle} {\emph {\bibinfo
  {booktitle} {{Proceedings: Workshop on the Unified Theories and the Baryon
  Number in the Universe: Tsukuba, Japan, February 13-14, 1979}}},\ }\href@noop
  {} {\bibfield  {journal} {\bibinfo  {journal} {Conf. Proc.}\ }\textbf
  {\bibinfo {volume} {C7902131}},\ \bibinfo {pages} {95} (\bibinfo {year}
  {1979})}\BibitemShut {NoStop}%
\bibitem [{\citenamefont {Gell-Mann}\ \emph {et~al.}(1979)\citenamefont
  {Gell-Mann}, \citenamefont {Ramond},\ and\ \citenamefont
  {Slansky}}]{GellMann:1980vs}%
  \BibitemOpen
  \bibfield  {author} {\bibinfo {author} {\bibfnamefont {M.}~\bibnamefont
  {Gell-Mann}}, \bibinfo {author} {\bibfnamefont {P.}~\bibnamefont {Ramond}},\
  and\ \bibinfo {author} {\bibfnamefont {R.}~\bibnamefont {Slansky}},\
  }\bibfield  {title} {\bibinfo {title} {{Complex Spinors and Unified
  Theories}},\ }\bibfield  {booktitle} {\emph {\bibinfo {booktitle}
  {{Supergravity Workshop Stony Brook, New York, September 27-28, 1979}}},\
  }\href@noop {} {\bibfield  {journal} {\bibinfo  {journal} {Conf. Proc.}\
  }\textbf {\bibinfo {volume} {C790927}},\ \bibinfo {pages} {315} (\bibinfo
  {year} {1979})},\ \Eprint {https://arxiv.org/abs/1306.4669} {arXiv:1306.4669
  [hep-th]} \BibitemShut {NoStop}%
\bibitem [{\citenamefont {Esteban}\ \emph {et~al.}(2019)\citenamefont
  {Esteban}, \citenamefont {Gonzalez-Garcia}, \citenamefont
  {Hernandez-Cabezudo}, \citenamefont {Maltoni},\ and\ \citenamefont
  {Schwetz}}]{Esteban:2018azc}%
  \BibitemOpen
  \bibfield  {author} {\bibinfo {author} {\bibfnamefont {I.}~\bibnamefont
  {Esteban}}, \bibinfo {author} {\bibfnamefont {M.~C.}\ \bibnamefont
  {Gonzalez-Garcia}}, \bibinfo {author} {\bibfnamefont {A.}~\bibnamefont
  {Hernandez-Cabezudo}}, \bibinfo {author} {\bibfnamefont {M.}~\bibnamefont
  {Maltoni}},\ and\ \bibinfo {author} {\bibfnamefont {T.}~\bibnamefont
  {Schwetz}},\ }\bibfield  {title} {\bibinfo {title} {{Global analysis of
  three-flavour neutrino oscillations: synergies and tensions in the
  determination of $\theta_23, \delta_CP$, and the mass ordering}},\ }\href
  {https://doi.org/10.1007/JHEP01(2019)106} {\bibfield  {journal} {\bibinfo
  {journal} {JHEP}\ }\textbf {\bibinfo {volume} {01}},\ \bibinfo {pages}
  {106}},\ \Eprint {https://arxiv.org/abs/1811.05487} {arXiv:1811.05487
  [hep-ph]} \BibitemShut {NoStop}%
\bibitem [{\citenamefont {Baldini}\ \emph {et~al.}(2016)\citenamefont {Baldini}
  \emph {et~al.}}]{TheMEG:2016wtm}%
  \BibitemOpen
  \bibfield  {author} {\bibinfo {author} {\bibfnamefont {A.~M.}\ \bibnamefont
  {Baldini}} \emph {et~al.} (\bibinfo {collaboration} {MEG}),\ }\bibfield
  {title} {\bibinfo {title} {{Search for the lepton flavour violating decay
  $\mu ^+ \rightarrow \mathrm {e}^+ \gamma $ with the full dataset of the MEG
  experiment}},\ }\href {https://doi.org/10.1140/epjc/s10052-016-4271-x}
  {\bibfield  {journal} {\bibinfo  {journal} {Eur. Phys. J.}\ }\textbf
  {\bibinfo {volume} {C76}},\ \bibinfo {pages} {434} (\bibinfo {year}
  {2016})},\ \Eprint {https://arxiv.org/abs/1605.05081} {arXiv:1605.05081
  [hep-ex]} \BibitemShut {NoStop}%
\bibitem [{\citenamefont {Kitano}\ and\ \citenamefont
  {Nomura}(2005)}]{Kitano:2005wc}%
  \BibitemOpen
  \bibfield  {author} {\bibinfo {author} {\bibfnamefont {R.}~\bibnamefont
  {Kitano}}\ and\ \bibinfo {author} {\bibfnamefont {Y.}~\bibnamefont
  {Nomura}},\ }\bibfield  {title} {\bibinfo {title} {{A Solution to the
  supersymmetric fine-tuning problem within the MSSM}},\ }\href
  {https://doi.org/10.1016/j.physletb.2005.10.003} {\bibfield  {journal}
  {\bibinfo  {journal} {Phys. Lett.}\ }\textbf {\bibinfo {volume} {B631}},\
  \bibinfo {pages} {58} (\bibinfo {year} {2005})},\ \Eprint
  {https://arxiv.org/abs/hep-ph/0509039} {arXiv:hep-ph/0509039 [hep-ph]}
  \BibitemShut {NoStop}%
\bibitem [{\citenamefont {Choi}\ \emph {et~al.}(2007)\citenamefont {Choi},
  \citenamefont {Jeong}, \citenamefont {Kobayashi},\ and\ \citenamefont
  {Okumura}}]{Choi:2006xb}%
  \BibitemOpen
  \bibfield  {author} {\bibinfo {author} {\bibfnamefont {K.}~\bibnamefont
  {Choi}}, \bibinfo {author} {\bibfnamefont {K.~S.}\ \bibnamefont {Jeong}},
  \bibinfo {author} {\bibfnamefont {T.}~\bibnamefont {Kobayashi}},\ and\
  \bibinfo {author} {\bibfnamefont {K.-i.}\ \bibnamefont {Okumura}},\
  }\bibfield  {title} {\bibinfo {title} {{TeV Scale Mirage Mediation and
  Natural Little SUSY Hierarchy}},\ }\href
  {https://doi.org/10.1103/PhysRevD.75.095012} {\bibfield  {journal} {\bibinfo
  {journal} {Phys. Rev.}\ }\textbf {\bibinfo {volume} {D75}},\ \bibinfo {pages}
  {095012} (\bibinfo {year} {2007})},\ \Eprint
  {https://arxiv.org/abs/hep-ph/0612258} {arXiv:hep-ph/0612258 [hep-ph]}
  \BibitemShut {NoStop}%
\bibitem [{\citenamefont {Abe}\ \emph {et~al.}(2007)\citenamefont {Abe},
  \citenamefont {Kim}, \citenamefont {Kobayashi},\ and\ \citenamefont
  {Shimizu}}]{Abe:2007je}%
  \BibitemOpen
  \bibfield  {author} {\bibinfo {author} {\bibfnamefont {H.}~\bibnamefont
  {Abe}}, \bibinfo {author} {\bibfnamefont {Y.~G.}\ \bibnamefont {Kim}},
  \bibinfo {author} {\bibfnamefont {T.}~\bibnamefont {Kobayashi}},\ and\
  \bibinfo {author} {\bibfnamefont {Y.}~\bibnamefont {Shimizu}},\ }\bibfield
  {title} {\bibinfo {title} {{TeV scale partial mirage unification and
  neutralino dark matter}},\ }\href
  {https://doi.org/10.1088/1126-6708/2007/09/107} {\bibfield  {journal}
  {\bibinfo  {journal} {JHEP}\ }\textbf {\bibinfo {volume} {09}},\ \bibinfo
  {pages} {107}},\ \Eprint {https://arxiv.org/abs/0706.4349} {arXiv:0706.4349
  [hep-ph]} \BibitemShut {NoStop}%
\bibitem [{\citenamefont {Kobayashi}\ \emph {et~al.}(2013)\citenamefont
  {Kobayashi}, \citenamefont {Makino}, \citenamefont {Okumura}, \citenamefont
  {Shimomura},\ and\ \citenamefont {Takahashi}}]{Kobayashi:2012ee}%
  \BibitemOpen
  \bibfield  {author} {\bibinfo {author} {\bibfnamefont {T.}~\bibnamefont
  {Kobayashi}}, \bibinfo {author} {\bibfnamefont {H.}~\bibnamefont {Makino}},
  \bibinfo {author} {\bibfnamefont {K.-i.}\ \bibnamefont {Okumura}}, \bibinfo
  {author} {\bibfnamefont {T.}~\bibnamefont {Shimomura}},\ and\ \bibinfo
  {author} {\bibfnamefont {T.}~\bibnamefont {Takahashi}},\ }\bibfield  {title}
  {\bibinfo {title} {{TeV scale mirage mediation in NMSSM}},\ }\href
  {https://doi.org/10.1007/JHEP01(2013)081} {\bibfield  {journal} {\bibinfo
  {journal} {JHEP}\ }\textbf {\bibinfo {volume} {01}},\ \bibinfo {pages}
  {081}},\ \Eprint {https://arxiv.org/abs/1204.3561} {arXiv:1204.3561 [hep-ph]}
  \BibitemShut {NoStop}%
\bibitem [{\citenamefont {Asano}\ and\ \citenamefont
  {Higaki}(2012)}]{Asano:2012sv}%
  \BibitemOpen
  \bibfield  {author} {\bibinfo {author} {\bibfnamefont {M.}~\bibnamefont
  {Asano}}\ and\ \bibinfo {author} {\bibfnamefont {T.}~\bibnamefont {Higaki}},\
  }\bibfield  {title} {\bibinfo {title} {{Natural supersymmetric spectrum in
  mirage mediation}},\ }\href {https://doi.org/10.1103/PhysRevD.86.035020}
  {\bibfield  {journal} {\bibinfo  {journal} {Phys. Rev.}\ }\textbf {\bibinfo
  {volume} {D86}},\ \bibinfo {pages} {035020} (\bibinfo {year} {2012})},\
  \Eprint {https://arxiv.org/abs/1204.0508} {arXiv:1204.0508 [hep-ph]}
  \BibitemShut {NoStop}%
\bibitem [{\citenamefont {Hagimoto}\ \emph {et~al.}(2016)\citenamefont
  {Hagimoto}, \citenamefont {Kobayashi}, \citenamefont {Makino}, \citenamefont
  {Okumura},\ and\ \citenamefont {Shimomura}}]{Hagimoto:2015tua}%
  \BibitemOpen
  \bibfield  {author} {\bibinfo {author} {\bibfnamefont {K.}~\bibnamefont
  {Hagimoto}}, \bibinfo {author} {\bibfnamefont {T.}~\bibnamefont {Kobayashi}},
  \bibinfo {author} {\bibfnamefont {H.}~\bibnamefont {Makino}}, \bibinfo
  {author} {\bibfnamefont {K.-i.}\ \bibnamefont {Okumura}},\ and\ \bibinfo
  {author} {\bibfnamefont {T.}~\bibnamefont {Shimomura}},\ }\bibfield  {title}
  {\bibinfo {title} {{Phenomenology of NMSSM in TeV scale mirage mediation}},\
  }\href {https://doi.org/10.1007/JHEP02(2016)089} {\bibfield  {journal}
  {\bibinfo  {journal} {JHEP}\ }\textbf {\bibinfo {volume} {02}},\ \bibinfo
  {pages} {089}},\ \Eprint {https://arxiv.org/abs/1509.05327} {arXiv:1509.05327
  [hep-ph]} \BibitemShut {NoStop}%
\bibitem [{\citenamefont {Kawamura}\ and\ \citenamefont
  {Omura}(2017)}]{Kawamura:2017qey}%
  \BibitemOpen
  \bibfield  {author} {\bibinfo {author} {\bibfnamefont {J.}~\bibnamefont
  {Kawamura}}\ and\ \bibinfo {author} {\bibfnamefont {Y.}~\bibnamefont
  {Omura}},\ }\bibfield  {title} {\bibinfo {title} {{Analysis of the TeV-scale
  mirage mediation with heavy superparticles}},\ }\href
  {https://doi.org/10.1007/JHEP11(2017)189} {\bibfield  {journal} {\bibinfo
  {journal} {JHEP}\ }\textbf {\bibinfo {volume} {11}},\ \bibinfo {pages}
  {189}},\ \Eprint {https://arxiv.org/abs/1710.03412} {arXiv:1710.03412
  [hep-ph]} \BibitemShut {NoStop}%
\bibitem [{\citenamefont {Pomarol}\ and\ \citenamefont
  {Rattazzi}(1999)}]{Pomarol:1999ie}%
  \BibitemOpen
  \bibfield  {author} {\bibinfo {author} {\bibfnamefont {A.}~\bibnamefont
  {Pomarol}}\ and\ \bibinfo {author} {\bibfnamefont {R.}~\bibnamefont
  {Rattazzi}},\ }\bibfield  {title} {\bibinfo {title} {{Sparticle masses from
  the superconformal anomaly}},\ }\href
  {https://doi.org/10.1088/1126-6708/1999/05/013} {\bibfield  {journal}
  {\bibinfo  {journal} {JHEP}\ }\textbf {\bibinfo {volume} {05}},\ \bibinfo
  {pages} {013}},\ \Eprint {https://arxiv.org/abs/hep-ph/9903448}
  {arXiv:hep-ph/9903448 [hep-ph]} \BibitemShut {NoStop}%
\bibitem [{\citenamefont {Rattazzi}\ \emph {et~al.}(2000)\citenamefont
  {Rattazzi}, \citenamefont {Strumia},\ and\ \citenamefont
  {Wells}}]{Rattazzi:1999qg}%
  \BibitemOpen
  \bibfield  {author} {\bibinfo {author} {\bibfnamefont {R.}~\bibnamefont
  {Rattazzi}}, \bibinfo {author} {\bibfnamefont {A.}~\bibnamefont {Strumia}},\
  and\ \bibinfo {author} {\bibfnamefont {J.~D.}\ \bibnamefont {Wells}},\
  }\bibfield  {title} {\bibinfo {title} {{Phenomenology of deflected anomaly
  mediation}},\ }\href {https://doi.org/10.1016/S0550-3213(00)00130-9}
  {\bibfield  {journal} {\bibinfo  {journal} {Nucl. Phys.}\ }\textbf {\bibinfo
  {volume} {B576}},\ \bibinfo {pages} {3} (\bibinfo {year} {2000})},\ \Eprint
  {https://arxiv.org/abs/hep-ph/9912390} {arXiv:hep-ph/9912390 [hep-ph]}
  \BibitemShut {NoStop}%
\bibitem [{\citenamefont {Okada}(2002)}]{Okada:2002mv}%
  \BibitemOpen
  \bibfield  {author} {\bibinfo {author} {\bibfnamefont {N.}~\bibnamefont
  {Okada}},\ }\bibfield  {title} {\bibinfo {title} {{Positively deflected
  anomaly mediation}},\ }\href {https://doi.org/10.1103/PhysRevD.65.115009}
  {\bibfield  {journal} {\bibinfo  {journal} {Phys. Rev.}\ }\textbf {\bibinfo
  {volume} {D65}},\ \bibinfo {pages} {115009} (\bibinfo {year} {2002})},\
  \Eprint {https://arxiv.org/abs/hep-ph/0202219} {arXiv:hep-ph/0202219
  [hep-ph]} \BibitemShut {NoStop}%
\bibitem [{\citenamefont {Everett}\ \emph
  {et~al.}(2008{\natexlab{a}})\citenamefont {Everett}, \citenamefont {Kim},
  \citenamefont {Ouyang},\ and\ \citenamefont {Zurek}}]{Everett:2008qy}%
  \BibitemOpen
  \bibfield  {author} {\bibinfo {author} {\bibfnamefont {L.~L.}\ \bibnamefont
  {Everett}}, \bibinfo {author} {\bibfnamefont {I.-W.}\ \bibnamefont {Kim}},
  \bibinfo {author} {\bibfnamefont {P.}~\bibnamefont {Ouyang}},\ and\ \bibinfo
  {author} {\bibfnamefont {K.~M.}\ \bibnamefont {Zurek}},\ }\bibfield  {title}
  {\bibinfo {title} {{Deflected Mirage Mediation: A Framework for Generalized
  Supersymmetry Breaking}},\ }\href
  {https://doi.org/10.1103/PhysRevLett.101.101803} {\bibfield  {journal}
  {\bibinfo  {journal} {Phys. Rev. Lett.}\ }\textbf {\bibinfo {volume} {101}},\
  \bibinfo {pages} {101803} (\bibinfo {year} {2008}{\natexlab{a}})},\ \Eprint
  {https://arxiv.org/abs/0804.0592} {arXiv:0804.0592 [hep-ph]} \BibitemShut
  {NoStop}%
\bibitem [{\citenamefont {Everett}\ \emph
  {et~al.}(2008{\natexlab{b}})\citenamefont {Everett}, \citenamefont {Kim},
  \citenamefont {Ouyang},\ and\ \citenamefont {Zurek}}]{Everett:2008ey}%
  \BibitemOpen
  \bibfield  {author} {\bibinfo {author} {\bibfnamefont {L.~L.}\ \bibnamefont
  {Everett}}, \bibinfo {author} {\bibfnamefont {I.-W.}\ \bibnamefont {Kim}},
  \bibinfo {author} {\bibfnamefont {P.}~\bibnamefont {Ouyang}},\ and\ \bibinfo
  {author} {\bibfnamefont {K.~M.}\ \bibnamefont {Zurek}},\ }\bibfield  {title}
  {\bibinfo {title} {{Moduli Stabilization and Supersymmetry Breaking in
  Deflected Mirage Mediation}},\ }\href
  {https://doi.org/10.1088/1126-6708/2008/08/102} {\bibfield  {journal}
  {\bibinfo  {journal} {JHEP}\ }\textbf {\bibinfo {volume} {08}},\ \bibinfo
  {pages} {102}},\ \Eprint {https://arxiv.org/abs/0806.2330} {arXiv:0806.2330
  [hep-ph]} \BibitemShut {NoStop}%
\bibitem [{\citenamefont {Choi}\ \emph {et~al.}(2009)\citenamefont {Choi},
  \citenamefont {Jeong}, \citenamefont {Nakamura}, \citenamefont {Okumura},\
  and\ \citenamefont {Yamaguchi}}]{Choi:2009jn}%
  \BibitemOpen
  \bibfield  {author} {\bibinfo {author} {\bibfnamefont {K.}~\bibnamefont
  {Choi}}, \bibinfo {author} {\bibfnamefont {K.~S.}\ \bibnamefont {Jeong}},
  \bibinfo {author} {\bibfnamefont {S.}~\bibnamefont {Nakamura}}, \bibinfo
  {author} {\bibfnamefont {K.-i.}\ \bibnamefont {Okumura}},\ and\ \bibinfo
  {author} {\bibfnamefont {M.}~\bibnamefont {Yamaguchi}},\ }\bibfield  {title}
  {\bibinfo {title} {{Sparticle masses in deflected mirage mediation}},\ }\href
  {https://doi.org/10.1088/1126-6708/2009/04/107} {\bibfield  {journal}
  {\bibinfo  {journal} {JHEP}\ }\textbf {\bibinfo {volume} {04}},\ \bibinfo
  {pages} {107}},\ \Eprint {https://arxiv.org/abs/0901.0052} {arXiv:0901.0052
  [hep-ph]} \BibitemShut {NoStop}%
\bibitem [{\citenamefont {Altunkaynak}\ \emph {et~al.}(2010)\citenamefont
  {Altunkaynak}, \citenamefont {Nelson}, \citenamefont {Everett}, \citenamefont
  {Kim},\ and\ \citenamefont {Rao}}]{Altunkaynak:2010xe}%
  \BibitemOpen
  \bibfield  {author} {\bibinfo {author} {\bibfnamefont {B.}~\bibnamefont
  {Altunkaynak}}, \bibinfo {author} {\bibfnamefont {B.~D.}\ \bibnamefont
  {Nelson}}, \bibinfo {author} {\bibfnamefont {L.~L.}\ \bibnamefont {Everett}},
  \bibinfo {author} {\bibfnamefont {I.-W.}\ \bibnamefont {Kim}},\ and\ \bibinfo
  {author} {\bibfnamefont {Y.}~\bibnamefont {Rao}},\ }\bibfield  {title}
  {\bibinfo {title} {{Phenomenological Implications of Deflected Mirage
  Mediation: Comparison with Mirage Mediation}},\ }\href
  {https://doi.org/10.1007/JHEP05(2010)054} {\bibfield  {journal} {\bibinfo
  {journal} {JHEP}\ }\textbf {\bibinfo {volume} {05}},\ \bibinfo {pages}
  {054}},\ \Eprint {https://arxiv.org/abs/1001.5261} {arXiv:1001.5261 [hep-ph]}
  \BibitemShut {NoStop}%
\end{thebibliography}%

\end{document}